\documentclass[graybox]{svmult}

\usepackage{natbib}
\usepackage{xspace}
\usepackage{filecontents}
\usepackage{graphicx}
\usepackage{amsmath}

\usepackage{listings}
\usepackage{mathabx}
\usepackage{url}

\usepackage{mathptmx}       
\usepackage{helvet}         
\usepackage{courier}        
\usepackage{type1cm}        
\usepackage{makeidx}         
\usepackage{graphicx}        
\usepackage{multicol}        
\usepackage[bottom]{footmisc}

\makeindex             

\newlength{\codeheadercommentindent}
\newcommand{\strong}[1]{}
\newcommand{\latin}[1]{{\it #1}}
\newcommand{\ie}{\latin{i.e.}\@\xspace}
\newcommand{\eg}{\latin{e.g.}\@\xspace}

\newcommand{\etc}{\latin{etc.}\@\xspace}

\newcommand{\ave}[1]{\left\langle #1 \right\rangle}

\newcommand{\abbrevdef}[2]{#1 (#2)}
\newcommand{\keyword}[1]{{\bf #1}}

\newcommand{\pdf}[2]{\mathcal{P}^{\supmarker{#1}}\spacetobracket\left(#2\right)}
\makeatletter
\newcommand{\supmarker}[1]{{\@ifempty{#1}{}{\text{(#1)}}}}
\makeatother
\newcommand{\spacetobracket}{\!\!}

\newcommand{\gpvec}[1]{\mathbf{#1}}

\newcommand{\nvec}{\gpvec{n}}

\usepackage{dsfont}

\newcommand{\gpset}[1]{\mathds{#1}}

\newcommand{\Rset}{\gpset{R}}

\newcommand{\GC}{\mathcal{G}}
\newcommand{\GCtilde}{\tilde{\GC}}
\newcommand{\gs}[1]{g_{#1}}

\newcommand{\plaind}{\mathrm{d}}

\newcommand{\dint}[1]{\mathchoice{\!\plaind#1\,}{\!\plaind#1\,}{\!\plaind#1\,}{\!\plaind#1\,}}

\newcommand{\ldf}{z}

\newcommand{\posscitet}[1]{\citeauthor{#1}'s (\citeyear{#1})}

\newcommand{\PG}[1]{#1}
\newcommand{\Opageref}[1]{{\PG{\pageref{#1}}}}
\newcommand{\plabel}[1]{\label{page:#1}}
\newcommand{\pref}[1]{\Pref{#1}}
\newcommand{\Pref}[1]{\PrefX{page:#1}}
\newcommand{\PrefX}[1]{p.~\Opageref{#1}}

\newcommand{\elabel}[1]{\label{eq:#1}}

\newcommand{\Eref}[1]{Eq.~(\ref{eq:#1})}

\newcommand{\slabel}[1]{\label{sec:#1}}
\newcommand{\sref}[1]{Sec.~\ref{sec:#1}}
\newcommand{\Sref}[1]{\sref{#1}}

\newcommand{\flabel}[1]{\label{fig:#1}}

\newcommand{\Fref}[1]{Figure~\ref{fig:#1}}

\newcommand{\code}[1]{\lstinline[breaklines=true,postbreak={},basicstyle=\ttfamily\small]$#1$}

\begin{document}
\title{SOC computer simulations}
\author{Gunnar Pruessner}
\institute{Gunnar Pruessner \at 
Imperial College London\\
Department of Mathematics\\
\email{g.pruessner@imperial.ac.uk}}

\maketitle

\abstract{
The following chapter provides an overview of the techniques used to
understand Self-Organised Criticality (SOC) by performing computer simulations.  Those are of
particular significance in SOC, given its very paradigm, the BTW
(Bak-Tang-Wiesenfeld)
sandpile, was introduced on the basis of a process that is conveniently
implemented as a computer program. The chapter is divided into three
sections: In the first section a number of key concepts are introduced,
followed by four  brief presentations of SOC models which are most
commonly investigated or which have played an important part in the
development of the field as a whole.  The second section is concerned
with the basics of scaling with particular emphasis of its r{\^o}le in
numerical models of SOC, introducing a number of basic tools for data
analysis such as binning, moment analysis and error estimation. The
third section is devoted to numerical methods and algorithms as applied
to SOC models, addressing typical computational questions with the
particular application of SOC in mind. The present chapter is rather
technical, but hands-on at the same time, providing practical advice and
even code snippets (in C) wherever possible.}


\section{Introduction}
The concept of \abbrevdef{Self-Organised Criticality}{SOC}\footnote{A
more extensive review on the present subject area can be found in
\citep{Pruessner:2012:Book}.}  was
introduced by \citet{BakTangWiesenfeld:1987} on the basis of a computer
model, the famous BTW Sandpile. The notion of ``computer model'' and
``simulation'' used here is subtle and can be misleading.  Often the
models are not meant to mimic a particular (natural) phenomenon, but are
intended to capture merely what is considered to be the \emph{essential}
interaction
observed in a natural phenomenon. Per Bak in particular, had the tendency to
name models according to their appearance rather than their purpose and
so the ``Sandpile Model'' may not have been envisaged to display the
dynamics of a sandpile. The situation is clearer in the case of the
``Forest Fire Model'' \citep{BakChenTang:1990}, which was developed as a
model of turbulence much more than
as a model of fires in woods.

In particular in the early days of SOC modelling, the models were
sometimes
referred to as ``cellular automata''
\cite{OlamiFederChristensen:1992,LebowitzMaesSpeer:1990}, which caused
some consternation \citep[\eg][]{Grassberger:1994}, as cellular automata
normally have discrete states and evolve in discrete time steps
according to deterministic rules in discrete space (\ie a lattice). The term ``coupled map lattice''
\citep{Kaneko:1989} can be more appropriate for some models, such as the
Olami-Feder-Christensen Model descusssed below (discrete space, continuous state and
possibly continuous time).

The terminology of ``numerical modelling'' has always been somewhat
confusing. Many of the models considered in SOC do not model a natural
phenomenon and so their numerical implementation is not a ``numerical
simulation'' in the sense that they mimic the behaviour of something
else. There are notable exceptions, however, such as the Forest Fire
Model \citep{BakChenTang:1990} mentioned above and the Oslo ricepile model
\citep{ChristensenETAL:1996}.
SOC models generally are not ``models of SOC'', rather they are
algorithmic prescriptions or ``recipes'' for a (stochastic) process that
is believed to exhibit some of the features normally observed in other
SOC models. In that sense, the terminology of terms like ``SOC models'' and
``simulation'' or even ``simulating an SOC model'' is misleading ---
most of these models are not simplified versions or idealisations of some physical
process or anything else that is readily identified as ``SOC'', but recipes
to produce some of the behaviour expected in an SOC system.

To this day, a large fraction of the SOC community dedicate their
research to computer models. Initially, the motivation
\citep[\eg][]{Zhang:1989,Manna:1991a} was to find
models displaying the same universal behaviour as the BTW
(Bak-Tang-Wiesenfeld) Sandpile. This was followed by an era of proliferation,
when many new models, belonging to new universality classes where
developed. More recently, in a more reductionistic spirit, new models
are mostly developed to isolate the r{\^o}le of particular features and
to extract and identify their effect \citep[\eg][]{TadicDhar:1997}. A
lot of numerical research into SOC nowadays happens ``en passant'', as
SOC is identified in a model for a phenomenon that originally was not
considered to be related to
SOC \citep[\eg][]{BurridgeKnopoff:1967}.

Virtually all SOC (computer) models consist of degrees of freedom interacting with
(nearest) neighbours located on a lattice. The degrees of freedom may be
parameterised by continuous or discrete variables, in the following denoted
$\ldf_{\nvec}$, where $\nvec$ is a position vector on the lattice.
A \keyword{slow, external driving
mechanism} (in short, \keyword{external drive}) slowly loads the system,
\ie the local variables are slowly increased, also referred to as
``charging a site''. That might happen
uniformly (sometimes called \keyword{global drive}) or at individual
lattice sites (sometimes called \keyword{point drive}). The driving
might happen at randomly chosen points or by random increments, both of
which is in the literature referred to as \keyword{random driving}. The
dynamics of an SOC model is \keyword{non-linear}, \ie there is no linear
equation of motion that would describe their dynamics.\footnote{It is
very instructive to ask why a non-linearity is such a crucial ingredient.
Firstly, if all interactions were linear, one would expect the resulting
behaviour to correspond to that of a solvable, ``trivial'' system.
Secondly, linearity suggests additivity of external drive and response,
so responses would be expected to be proportional to the drive, a rather
boring behaviour, not expected to result in scale invariance.} 
The response of the system is triggered by a local
degree of freedom overcoming a 
\keyword{threshold}, beyond which \keyword{relaxation} and thus interaction with other degrees of
freedom and the outside world takes place. A site where that happens is
said to \keyword{topple} and to be \keyword{active}. The interaction might lead to
one of the neighbours exceeding its threshold in turn, triggering
another relaxation event. The totality of the relaxations constitutes an
\keyword{avalanche}. When the avalanche has finished, \ie there are no 
active sites left, the system is in a state of \keyword{quiescence}.
In SOC models, driving takes place only in the quiescent state
(separation of time scales, below). If the
external drive acts at times when an avalanche is running, it might lead
to a continuously running avalanche \citep[e.g.][]{CorralPaczuski:1999}.

In many models the degree of freedom at every site measures a resource
that is \keyword{conserved} under the dynamics. To balance the external drive, in
most models \keyword{dissipation} has to take place in some form:
\keyword{Bulk dissipation} takes place when the resource can get lost
in the local interaction. \keyword{Boundary dissipation} refers to the
situation when the resource is lost only in case a boundary site
relaxes. The necessary flux of the resource towards the boundaries has
been suggested as some of the key mechanisms in SOC
\citep{PaczuskiBassler:2000}. In some models, such as the Bak-Sneppen
Model \citep{BakSneppen:1993} or the Forest-Fire-Models
\citep{Henley:1989,BakChenTang:1990,DrosselSchwabl:1992a}, no (limited) resource can be
identified and therefore the notion of dissipation and
conservation is not meaningful.

The question whether conservation is a necessary ingredient of SOC has
driven the evolution of SOC models in particular during the 1990s. In
fact, early theoretical results by \citet{HwaKardar:1989a} suggested
that bulk dissipation would spoil the SOC state. Models like the
OFC Model \citep[][also
\citealp{BakSneppen:1993,DrosselSchwabl:1992a}]{OlamiFederChristensen:1992}
questioned that finding. Different theoretical views have emerged over
time: \posscitet{LauritsenZapperiStanley:1996} self-organised branching
process \citep{ZapperiLauritsenStanley:1995} contains dissipation as a \emph{relevant} parameter which
has a limitting effect on the scaling behaviour. 
\citet{JuanicoMonterolaSaloma:2007b} restored the SOC state of the
self-organised branching process by implementing a mechanism
that compensates for the non-conservation by a ``matching condition'' not
dissimilar from the mechanism used in the mean-field theory by
\citet{JensenPruessner:2002a}. That, in turn, was labelled by
\citet{BonachelaMunoz:2009} as a form of tuning. More recent
field-theoretic work \citep{Pruessner:2012:FT} points at conservation as a symmetry responsible
for the cancellation of mass-generating diagrams, an effect that may
equally
be achieved by other symmetries.

The external drive, the ensuing
sequence of avalanches and the evolution of the model from one
quiescent state to the next happen on the \keyword{macroscopic time
scale}, where time typically passes by one unit per avalanche. As the
system size is increased, avalanches are expected to take more and more
relaxations to complete. Their duration is measured on the 
\keyword{microscopic time scale}. 
In the thermodynamic limit, \ie at
infinite system size, the infinite duration of an avalanche of the
microscopic time scale and the finite driving rate on the macroscopic
time scale amount to a complete \keyword{separation of time scales}.
In general, the separation of time scales is achieved in finite systems
provided that no driving takes place when any site is active, because
the times of quiescence, measured on the microscopic time scale, can be
thought of as arbitrarily long. As a result, the avalanching in these
systems becomes \keyword{intermittent}.

Separation of time scales is widely regarded as \emph{the} crucial
ingredient of SOC, maybe because it is conceived (and criticised as
such) as a substitute of the tuning found in traditional critical
phenomena \citep[also][]{Jensen:1998}.  In numerical models, it normally
enters in a rather innocent way --- the system is not driven while an
avalanche is running. This, however, requires some global supervision, a
``babysitter'' \citep{DickmanETAL:2000} or a ``farmer''
\citep{BrokerGrassberger:1999}. In some models the separation of time
scales can be implemented explicitly \citep{BakSneppen:1993} in the
relaxational rule.  What makes the separation of time scales very
different from other forms of tuning is that it \emph{eliminates} a
dimensionful, finite scale, such as the frequency with which an
avalanche is triggered.\footnote{In the field theory of SOC, the
cancellation of diagrams occurs precisely when stationarity is imposed
for the density of particles resting (and their correlations) in the
limit $\omega\to0$, \ie in the long time limit.} In traditional critical
phenomena, scaling comes about due to the \emph{presence} of a
dimensionful, finite energy scale\footnote{For example $k_B T_c$ in the
Ising Model \citep{Stanley:1971}.}, where entropic contributions to the free
energy compete with those from the internal energy promoting order. In
most SOC models, it is pretty obvious that scaling would break down if
time scales were not explicitly separated --- avalanches start merging
and eventually intermittency is no longer observed
\citep{CorralPaczuski:1999}.

SOC models are normally studied at \keyword{stationarity}, when all correlations
originating from the initial state (often the empty lattice) are negligible. Reaching
this point is a process
normally referred to as \keyword{equilibration}. The
equilibration time is normally measured as the number of charges by the
external drive required to reach stationarity.
For some
models, exact upper bounds for the equilibration time are known
\citep[\eg]{DharETAL:1995,Corral:2004c,Dhar:2004}. In deterministic
models, a clear distinction exists between \keyword{transient} and
\keyword{recurrent
states}, where the former can appear at most once, and the latter with a
finite frequency provided the number of states overall is finite. In
fact, this frequency is the same for all recurrent states, depending on
the driving, which can be at one site only or randomly and independently
throughout. A detailed proof of such properties can be cumbersome
\citep{Dhar:1999a,Dhar:1999c}.

The statistics of the avalanches, their size as
well as
their extent in space and in time, is collected and analysed. SOC is
usually said to be found in these models when the statistics displays a
\keyword{scaling symmetry}, governed by only one upper cutoff which diverges with
the system size. In principle, a Gaussian possesses this scaling
symmetry,\footnote{The basic example $\pdf{}{s}=s^{-1} \GC(s/s_c)$ with
$\GC(x)=2 x \exp(-x^2)/\sqrt{\pi}$ is normalised and has avalanche size
exponent $\tau=1$, as defined in \Eref{simple_scaling}. \emph{Without}
the pre-factor $x$ in $\GC(x)$ the graph looks surprisingly similar to
a PDF as typically found in SOC models.}
but not a single important SOC model has a Gaussian
event size distribution.
On the contrary, the avalanche statistics of all models discussed below
deviates dramatically from a Gaussian, thus suggesting that avalanches
are not the result of essentially independent patches of avalanching
sites creating a bigger overall avalanche. Rather, sites are \keyword{strongly
interacting}, thereby creating the overall event. The purpose of
numerical simulations is to characterise and quantify this interaction
and its effect, as well as extracting \keyword{universal quantities}, which can be
compared with those found in other systems.

\subsection{Observables}
As for the methods of analysis, they have matured considerably over the
past decades. The initial hunt for $1/f$ noise in temporal signals has
given way to the study of event size distributions. As a matter of
numerical convenience, these distributions are often characterised using
moments, some of which are known exactly. Since the beginning of
computational physics, moments and
cumulants have been the commonly used method of choice to
characterise critical phenomena \citep{BinderHeermann:1997}. It is probably
owed to the time of the late 1980's that memory-intensive observables
such as entire distributions became computationally affordable and
subsequently the centre of attention in SOC.  

To this day, the analysis of moments in SOC is still often regarded as an
unfortunate necessity to characterise distributions, which are difficult
to describe quantitatively. Apart from the historic explanation alluded to
above, there is another, physical reason for that, the \keyword{avalanche size
exponent} $\tau$. In traditional critical phenomena, the corresponding
exponent of the order parameter distribution is
fixed at unity in the presence of the Rushbrooke and the Josephson
scaling law \citep{ChristensenETAL:2008}. The deviation of $\tau$ from
unity, which implies that the expected event size does not scale like
the characteristic event size, is another distinctive feature of SOC.
To some extent,
the exponent $\tau$ can be extracted from the avalanche size
distribution (almost) by inspection. In a moment analysis, on the other
hand, it is somewhat
``hidden'' in the details.

The most important observables usually extracted from an SOC model are
thus the scaling exponents, such as $\tau$, $D$ (\keyword{avalanche
dimension}), $\alpha$ (\keyword{avalanche duration exponent}) and $z$
(\keyword{dynamical exponent})
discussed below. Here, the two exponents $D$ and $z$ are generally
regarded as more universal than $\tau$ and $\alpha$, as the former is
often ``enslaved'' by an exact scaling law related to the average avalanche
size, and the latter by a similar scaling law based on the ``narrow
joint distribution assumption'', discussed in \Sref{scaling}. Generally,
all observables that are universal or suspected to be are of interest.
This includes the scaling function (\Sref{scaling}) which is most easily
characterised by moment ratios, corresponding to universal amplitude
ratios, traditionally studied in equilibrium critical phenomena
\citep{PrivmanHohenbergAharony:1991,SalasSokal:1999}.

\subsection{Models}
There is wide consensus on
a number of general features of SOC models which
seem to play a r{\^o}le in determining the universality class each
belongs to. The very first SOC model, the BTW model, was essentially
\keyword{deterministic}, \ie there was no randomness in the bulk relaxation. A
given configuration plus the site being charged next determines the
resulting configuration uniquely. Even in these models, however, there
can be a degree of \keyword{stochasticity}, namely when the site to be charged by
the external drive is chosen at random. Finally, even when this is not
the case, \ie external drive and internal relaxation are deterministic,
initial conditions are often chosen at random and averaged over.

Deterministic SOC models have the great appeal that they are
``autonomous'' (in a non-technical sense) or ``self-sufficient'' in
that they do not require an additional source of (uncorrelated) noise.
It is difficult to justify the existence of an external source which
produces white, Gaussian noise, as that noise correlator,
$\ave{\eta(t)\eta(t')}=2 \Gamma^2 \delta(t-t')$, itself displays a form
of scaling $\ave{\eta(\alpha t)\eta(\alpha t')}=\alpha^{-1}
\ave{\eta(t)\eta(t')}$. 
The presence of an external (scaling) noise source seems to demote an SOC model to a
conversion mechanism of scale invariance, which becomes most apparent
when the respective model is cast in the language of stochastic
equations of motion, \ie\,\keyword{Langevin equations}.

Famous examples of deterministic SOC models, which do not require an
external noise source for the relaxation process,
are the BTW model with deterministic drive \citep[][but
\citealp{Creutz:2004}]{BakTangWiesenfeld:1987}, the OFC model
\citep{OlamiFederChristensen:1992} and, closely related, the train model
\citep{deSousaVieira:1992}. Of these only the latter has been studied
extensively in the absence of all stochasticity.

Most SOC models, however, have a strong stochastic component, \ie there is some
randomness in the relaxation mechanism that gives rise to avalanches. In
fact, models with some form of built-in randomness seem to give cleaner
scaling behaviour, suggesting that deterministic models get ``stuck'' on
some trajectory on phase space, where some conservation law prevents
them from exploring the rest of phase space
\citep{BagnoliETAL:2003,CasartelliETAL:2006}. Notably, randomising the BTW
model seems to push it into the Manna universality class
\citep{KarmakarMannaStella:2005}. The latter model is probably the
simplest SOC model displaying the most robust and universal scaling
behaviour \citep{HuynhPruessnerChew:2011}. Due to
the noise, trajectories of particles deposited by the external drive are
those of random walkers.

The second dividing line distinguishes \keyword{Abelian} and
\keyword{non-Abelian} models. The term was coined by \citet{Dhar:1990a}
introducing, strictly speaking, the Abelian Sandpile Model, by
re-expressing the original BTW Model \citep{BakTangWiesenfeld:1987} in
terms of units of slope rather than local particle numbers. This convenient
choice of driving and boundary conditions renders the model unphysical
as entire rows of particles are added and removed at once. At the same
time, however, the model's final state after two consecutive charges at
two different sites becomes independent from the order in which the
charges and the subsequent relaxations are carried out. Practically all
analytical insight into the BTW model is based on \posscitet{Dhar:1990a}
Abelian version. Because it is easier to implement, it has also favoured
in numerical simulations.

The term ``Abelian'' seems to suggest the existence of a (commutative) group, \ie
a set of operators closed under consecutive application, associative and
containing inverse and an identity. For most SOC models referred to as
Abelian, no such group is known, for example because operators do not exist
explicitly, or the associative property makes little sense, similarly for the
identity. Crucially, inverse operators rarely exist. To label a model
Abelian therefore normally means that the final state does not depend on
the order in which external charges are applied, \ie the model updating
operators (whether or not they exist), which drive it at various
locations,
commute. Because the final state
is unique only in the case of deterministic models, stochastic models
are Abelian provided that the statistics of the final state does not
depend on the order in which external charges are applied
\citep{Dhar:1999c}. The operators, which generally depend on the site
the driving is applied to,
of deterministic models apply to a model's state and take it from one
quiescent state to the next. The operators in a stochastic model act on
the distribution of states, \ie they are the Markov operators. A
deterministic model can be cast in the same language, however, the
Markov operators then correspond to simple permutation matrices.

While Abelianness originally refers to the evolution of a model on the
\plabel{microscopic_Abelianness}
macroscopic time scale, it is generally used to characterise its
behaviour on the microscopic timescale, \ie the step-by-step,
toppling-to-toppling update. It is therefore usually concluded that the
properties of avalanches and their statistics is independent from the
order of toppling of multiple active sites.

Strictly, however, the Abelian symmetry does not apply to the
microscopic time scale, at least for two reasons. Firstly, the
Abelian operators apply, a priori, only to the avalanche-to-avalanche
evolution, \ie the macroscopic time scale.
What is more, they apply to the final state and its statistics, but not
necessarily to the observables. Applying charges at two different sites
of an Abelian SOC model, starting from the same configuration, results
in the same final state (or its statistics) regardless of the order in which
the charges were applied, but not necessarily in the same pair of avalanche
sizes produced. On the basis of the proof of Abelianness, at least in
deterministic models, this limitation
is alleviated by the insight that the sum of the avalanche
sizes is invariant under a change of the order in which the model is
charged.

As for the second reason, many models come with a detailed prescription of the
microscopic updating procedure and therefore the microscopic time scale.
Strictly, the invariance under a change of order of updates on the
microscopic time scale thus applies to different models. The situation
corresponds to equating different dynamics in the Ising model: For some
observables, Glauber 
dynamics is different from Heat Bath dynamics, yet both certainly
produce the same critical behaviour. In fact, choosing different dynamics
(and thereby possibly introducing new conserved symmetries) can lead to
different dynamical critical behaviour.

Revisting the proof of Abelianness, however, generally reveals that the
caveats above are overcautious. The very proof of Abelianness on the
macroscopic time scale uses and develops a notion of Abelianness on the
microscopic time scale. This connection can be made more formally, once
it has been established that any configuration, quiescent or not, can be
expressed by applying a suitable number of external charges on each site
of an empty lattice.

Abelianness generally plays a major r{\^o}le in the analytical treatment
of SOC models, because it allows significant algebraic simplifications,
not least when the dynamics of a model is written in terms of Markov
matrices. It applies, generally, equally to recurrent and transient
states, where no inverse exists.
It remains highly desirable to
demonstrate Abelianness on the basis of the algebra, once that is
established as a suitable representation of a model's dynamics.

In the following section a few paradigmatic models of SOC are
introduced: The BTW Model, the Manna Model, the OFC Model and the Forest
Fire Model.

\subsubsection{The BTW Model}
The BTW Model was introduced together with the very concept of SOC
\citep{BakTangWiesenfeld:1987}, initially to explain the ``ubiquity'' of
$1/f$ noise. Of course, since then, SOC has been studied very much in its
own right. Like virtually all SOC models, the BTW Model consists of a
set of rules that prescribe how a local degree of freedom $z_i$ on a
$d$-dimensional lattice with sites $i$ is to be updated. There are two different
stages, namely the relaxation and the driving, the latter considered to
be slow compared to the relaxation, \ie the relaxation generally is
instantaneous and never occurs simultaneously with the driving
(separation of time scales). In the
Abelian version of the BTW Model \citep{Dhar:1990a}, the driving consists
of adding a single slope unit \citep{KadanoffETAL:1989} to a site, that
is normally picked uniformly and at random. The lattice is often
initialised with $z_i=0$ for all $i$.

If the driving leads to any of the $z_i$ exceeding the 
the critical slope $z^c$ (also referred to as the critical height or
threshold, depending on the view)
at a site $i$
a \keyword{toppling} occurs whereby $z_i$ is reduced by
the coordination number $q$ of the site and $z_j$ of every nearest neighbour $j$ increases by one (sometimes
referred to as \keyword{charging}). In principle both $q$ and $z^c$ can
vary from site to site and such generalisations are trivial to
implement. It is common to choose $z^c=q-1$.

The rules of the BTW Model can be summarised as follows:
\begin{description}
\item[{\bf Initialisation:}] All sites $i$ are empty, $z_i=0$.
\item[{\bf Driving:}] One unit is added at a randomly chosen (or sometimes
fixed) site $i$, \ie $z_i \to z_i + 1$.
\item[{\bf Toppling:}] A site with $z_i>z^c=q-1$ (called \emph{active}) distributes one unit to
the $q$ nearest neighbouring sites $j$, so that $z_i\to z_i-q$ and
$z_j\to z_j+1$. 
\item[{\bf Dissipation:}] Units are lost at boundaries, where
toppling site $i$ loses $q$ units, $z_i\to z_i-q$, yet less than $q$
nearest neighbours exist, which receive a unit.
\item[{\bf Time progression:}] Time progresses by one unit per parallel
update, when all active sites are
updated at once.
\end{description}

A toppling can trigger an avalanche, as charged neighbours might
exceed the threshold in turn, possibly by more than one unit. Strictly, the BTW Model is
updated in parallel, all sites topple at once whose local degree of
freedom exceeds
the threshold at the beginning of a time step.
Microscopic time
then advances by one unit. This way, $z_i$ might increase far beyond
$z^c$
before toppling itself. As long as $z_i>z^c$ for any site $i$, the sites in
the model carry on toppling. The totality of the toppling events is an
avalanche. 
In 
the Abelian BTW model as refined by \citet{Dhar:1990a}, the final state
of the model does not depend on the order in which external charges are
applied. In the process of the proof of this property, it turns out
that the order of processing any charges during the course of an
avalanche neither affects the final state nor the size of the avalanche
triggered. Using a parallel updating scheme or not therefore does not
change the avalanche sizes recorded.
As the order of updates defines the microscopic time scale, a change in
the updating procedure, however, affects all observables dependent on
that time, such as avalanche duration or correlations on the fast time
scale.

To keep
the prescription above consistent with the notion of boundary sites,
where toppling particles are to be lost to the outside, boundary sites have
to be thought of as having the same number of nearest neighbours as any
other, equivalent site in the bulk, except that some of their neighbours
are not capable of toppling themselves.
For numerical purposes it is
often advisable to embed a lattice in some ``padding'' (a
neighbourhood's ``halo'', see \Sref{neighbourhood_information},
\pref{padding}), \ie sites that
cannot topple but are otherwise identical to all other sites.

The sum of the slope units residing on a given site $i$ and those residing
on its nearest neighbours remains unchanged by the toppling of site $i$,
\ie the bulk dynamics in the BTW are conservative. Dissipation occurs
exclusively at the boundary and every slope unit added to the system in
the bulk must be transported to the boundary in order to leave the
system.

The original version of the BTW model is defined in terms of local
heights, so that the height differences give rise to the slope $z_i$,
which has to reach $q$ in order to trigger an a toppling. While this is
a perfectly isomorphic view of the BTW, \emph{driving} it in terms of
height units has a number of unwanted implications. In particular, it
loses its Abelianness. For that reason, the original version of the BTW
is rarely studied numerically nowadays.

The BTW Model is \keyword{deterministic} apart from the driving, which can be made
deterministic as well, simply by fixing the site that receives the
external charge that triggers the next avalanche. Even when slope units do not
move independently at toppling, a randomly chosen slope
unit being transported through a BTW system describes the trajectory of
a random
walker trajectories \citep{Dhar:1990a}, essentially because every
possible path is being realised (just not independently, but all with the
correct weight).
As a result, the average avalanche size $\ave{s}$ can be calculated exactly; The
number of moves a slope unit makes on average from the time of being added 
by the external drive to the time it leaves the system through an open
boundary is equal to the expected number of charges it causes. The
expected number of charges (caused by the movement of all slope units
taking part in an avalanche) per slope unit added is thus exactly equal to
the expected number of moves a slope unit makes until it leaves the
system, \ie its escape time. If the avalanche size is measured by the
number of topplings, which is more common, the expected number of moves
has to be divided by the number of moves per toppling, $q$ in the
present case. Higher moments of the avalanche size, or, say, the
avalanche size conditional to non-zero size (\ie at least one site
toppling in every avalanche), cannot be determined using the random walker approach, as
they are crucially dependent on the interaction of toppling sites.

Due to the random walker property of the slope units added, the scaling
of the average avalanche size thus merely depends on the particularities
of the driving. If the driving is random and uniform, then
$\ave{s}\propto L^2$ for any $d$-dimensional hypercubic lattice and 
\citep{RuelleSen:1992}
\begin{equation}
\ave{s}=\frac{1}{12} (L+1)(L+2) 
\elabel{exact_aves}
\end{equation}
in one dimension with two open boundaries, where the avalanche size is
the number of topplings per particle added. 
However, the dynamics of the BTW
Model in one dimension is trivial, so that the model is usually
studied only in $d=2$ and beyond. 

Because (or despite of) its deterministic nature, a large
number of analytical results are known, in one dimension
\citep{RuelleSen:1992} but more importantly in two dimensions
\citep{MajumdarDhar:1992}, not least on
the basis of (logarithmic) conformal field theory
\citep[e.g.][]{MajumdarDhar:1992,Ivashkevich:1994,MahieuRuelle:2001,Ruelle:2002,Jeng:2005b}.
Unfortunately, to this day, the scaling of the avalanche size
distribution in dimensions $d\ge2$ remains somewhat unclear.
Numerically, results are inconclusive, as different authors quote
widely varying results for $d=2$
\citep[e.g.]{VespignaniZapperiPietronero:1995,ChessaETAL:1999,LinHu:2002,Bonachela:2008},
possibly due to logarithmic corrections
\citep{Manna:1990,LuebeckUsadel:1997a,Luebeck:2000}

A major insight into the \emph{collective} dynamics of toppling sites
was the decomposition of avalanches into \keyword{waves}
\citep{IvashkevichKtitarevPriezzhev:1994a}, which was later used by
\citet{PriezzhevKtitarevIvashkevich:1996} to conjecture $\tau=6/5$ for
the avalanche size exponent in two dimensions. 
No site in an avalanche can topple more often than the site at which the
avalanche was triggered. Not allowing that first site to topple
therefore stops the avalanche from progressing any further and
each toppling of the first site thus defines a wave
of toppling.

While the BTW Model has been crucial for the formation of the field of
SOC as a whole, its poor convergence beyond one dimension has made it
fall in popularity. One may argue that the determinism of the
dynamics is to blame, as found in other models \citep{MiddletonTang:1995}. Indeed, adding some
stochasticity makes the BTW Model display the universal behaviour of the
Manna Model discussed in the next section \citep{Cernak:2002,Cernak:2006}. 

The exponents reported for the BTW Model vary greatly. In two
dimensions, the value of $\tau$ found in various studies ranges from $1$
\citep{BakTangWiesenfeld:1987} to $1.367$ \citep{LinHu:2002} and that
for $D$
from $2.50(5)$ \citep{DeMenechStellaTebaldi:1998} to $2.73(2)$
\citep{ChessaETAL:1999}. Similarly $\alpha$ is reported from $1.16(3)$
\citep{Bonachela:2008} to $1.480(11)$ \citep{LuebeckUsadel:1997a} and
$z$ from $1.02(5)$ \citep{DeMenechStella:2000} to $1.52(2)$
\citep{ChessaETAL:1999}. Using comparatively large system sizes,
\citet{DornHughesChristensen:2001} found exponents that seem to vary
systematically with the system size with little or no chance to identify
an asymptotic value.

The first exactly solved SOC model was the Dhar-Ramaswamy Model
\citep{DharRamaswamy:1989} which is the \keyword{directed} variant of the
BTW Model. The directedness means that during an individual avalanche,
sites are never re-visted, which effectively suppresses spatial
correlations. Random drive of the model results in a product state,
where sites taking part in an avalanche form a ``compact'' patch (\ie
they have no holes), which is delimited by boundaries describing a
random walk. The exponents in $d=d_{\perp}+1$ dimensions are given
analytically by $D=1+d_{\perp}/2$, $D(2-\tau)=1$, $z=1$ and
$D(\tau-1)=z(\alpha-1)$, which implies $\alpha=D$ and $\tau=2-1/D$
\citep{DharRamaswamy:1989,Christensen:1992,ChristensenOlami:1993,TadicDhar:1997,KlosterMaslovTang:2001}.
For example, in $d=1+1$ dimensions (directed square lattice), exponents
are $D=3/2$, $\tau=4/3$, $z=1$ and $\alpha=3/2$. Mean-field exponents
apply at $d=2+1$ and above.

\subsubsection{The Manna Model}
The \citet{Manna:1991a} Model was originally intended as a simplified
version of the BTW Model but has since then acquired the status of the
paradigmatic representative of the largest (and maybe the only)
universality class in SOC, generally refered to as the Manna, Oslo
\citep{ChristensenETAL:1996} or
C-DP \citep[conserved directed
percolation, ][]{RossiPastor-SatorrasVespignani:2000} universality class.

The Manna Model displays robust, clean critical behaviour
in any dimension $d\ge1$, characterised by non-trivial exponents below $d=4$
\citep{LuebeckHeger:2003b}. Originally, it is defined as follows: The
external drive adds particles at random chosen sites $i$, \ie the local
degree of freedom increases by one, $z_i\to z_i+1$. If a site exceeds
the threshold of $z^c=1$ it topples, so that \emph{all} its particles are redistributed
to the nearest neighbours, which are chosen independently at random.
After the toppling of site $i$, the local degree of freedom is therefore
set to $z_i=0$, while the total increase of the $z_j$ at the nearest
neighbours $j$ of $i$ maintains conservation. Again, as in the BTW
model, non-conservation at boundary sites can be thought of as been
implemented by sites that never topple themselves. 

Charging neighbours might push their local degree of freedom beyond the
threshold and they might therefore topple in turn. When a site topples,
all particles present there at the time of toppling are transferred to
its neighbour (maybe to a single one) and it is
therefore crucial to maintain the order of (parallel) updates. The model
is thus non-Abelian. In fact, the notion of Abelianness was initially restricted
to deterministic models \citep{MilshteinBihamSolomon:1998}. However,
\citet{Dhar:1999a} introduced a version of the Manna Model which is
Abelian in the sense that the statistics of the final state remains
unchanged if two consecutive external charges (by the driving) are
carried out in reverse order. In that version of the Manna Model, a
toppling site redistributes only $2$ of its particles, \ie the number of
particles redistributed at a toppling does not depend on $z_i$ itself.
The difference between the BTW Model and the Manna Model lies thus merely
in the fact that only two particles are re-distributed when a site
topples in the Manna Model (irrespective of the coordination number of
the site) and that the receiving sites are are picked
at random.

In summary, the rules of the Abelian Manna Model are:
\begin{description}
\item[{\bf Initialisation:}] All sites $i$ are empty, $z_i=0$.
\item[{\bf Driving:}] One unit is added at a randomly chosen (or sometimes
fixed) site $i$, \ie $z_i \to z_i + 1$.
\item[{\bf Toppling:}] A site with $z_i>z^c=1$ (called \emph{active}) distributes one unit to
$2$ randomly and independently chosen nearest neighbouring sites $j$, so that $z_i\to z_i-2$ and
$z_j\to z_j+1$. 
\item[{\bf Dissipation:}] Units are lost at boundaries, where the
randomly chosen nearest neighbour might be outside the system.
\item[{\bf Time progression:}] Originally, time progresses by one unit per parallel
update, when all active sites are
updated at once.
\end{description}

That the scaling in one dimension is not as clean as in higher dimension
may be caused by logarithmic corrections \citep{DickmanCampelo:2003}.
Nevertheless, it has been possible to extract consistent estimates for
exponents in dimensions $d=1$ to $d=5$
\citep{LuebeckHeger:2003b,HuynhPruessnerChew:2011,HuynhPruessner:2012b}.
Because some of its exponents are so similar to that of the directed
percolation 
universality class
\citep{Janssen:1981,Grassberger:1982,Hinrichsen:2000a}
there remains some doubt whether the Manna Model really represents a
universality class in its own right \citep{MunozETAL:1999,DickmanTomedeOliveira:2002}.
The problem is more pressing in the \keyword{fixed energy} version
\citep{DickmanVespignaniZapperi:1998,VespignaniETAL:1998} of the
Manna Model \citep{BasuETAL:2012}, where dissipation at boundaries is
switched off by closing them periodically, thereby studying the model at
a fixed amount of particles. The term ``fixed energy sandpile'' was
coined to stress the conserved nature of the relavent degree of freedom
(which may be called ``energy'') and to suggest a similar distinction
as in the change of ensemble from canonical to microcanonical.
\citet{BonachelaMunoz:2007}
suggested to study the model with different boundary conditions which
have an impact on the Manna Model that is distinctly different from
that on models in the directed percolation universality class.

Because of its fixed energy version, the Manna Model is frequently
studied for its links to absorbing state (AS) phase transitions
\citep{DickmanVespignaniZapperi:1998,VespignaniETAL:1998,Hinrichsen:2000a,HenkelHinrichsenLuebeck:2008}. In fact, it has
been suggested that SOC is due to the self-organisation to the critical
point of such an AS phase transition
\citep{TangBak:1988a,DickmanVespignaniZapperi:1998,VespignaniETAL:1998},
whereby strong activity leads to a reduction of particles by dissipation, making the
system in-active, while quiescence leads to activity due to the external
drive. One may argue that such a linear mechanism cannot produce the
desired universal critical behaviour without finely tuning the relevant
parameters
\citep{PetersPruessner:2006,PetersPruessner:2008,AlavaETAL:2008}.

A number of theoretical results are available for the Manna Model
\citep{VespignaniETAL:1998,VespignaniETAL:2000,RossiPastor-SatorrasVespignani:2000,vanWijland:2002,RamascoMunozdaSilvaSantos:2004},
yet an $\epsilon$-expansion \citep{LeDoussalWieseChauve:2002} for the Manna universality class is
available only via the mapping \citep{PaczuskiBoettcher:1996,Pruessner:2003} of the Oslo Model
\citep{ChristensenETAL:1996}, which is the same universality class
\citep{NakanishiSneppen:1997} as the Manna Model, to the quenched
Edwards-Wilkinson equation
\citep{BruinsmaAeppli:1984,KoplikLevine:1985,NattermannETAL:1992,LeschhornETAL:1997}.
Quenched noise and disorder are, however, notoriously difficult to
handle analytically.
It is thus highly desirable to develop a better theoretical
understanding of the Manna Model in its own right, including its
mechanism of self-organisation, and to derive an $\epsilon$-expansion
for its exponents.

Although the Manna Model is more frequently studied in one dimension,
for comparison with the BTW Model above, the exponents listed in the
following were determined numerically in two dimensions for the Abelian
and the non-Abelian (original) variant of the Manna Model. For $\tau$ they range
from $1.25(2)$ \citep{BihamMilshteinMalcai:2001} to $1.28(2)$
\citep{Manna:1991a,LuebeckHeger:2003a}, for $D$ from $2.54$
\citep{Ben-HurBiham:1996} to $2.764(10)$ \citep{Luebeck:2000}, for
$\alpha$ from $1.47(10)$ \citep{Manna:1991a} to $1.50(3)$
\citep{ChessaVespignaniZapperi:1999,LuebeckHeger:2003a} and for $z$ from
$1.49$ \citep{Ben-HurBiham:1996} to $1.57(4)$
\citep{AlavaMunoz:2002,DickmanTomedeOliveira:2002}, generally much more
consistent than in the BTW Model.

As in the BTW Model, various directed variants of the Manna Model
which are exactly solvable for similar reasons as in the deterministic
case have been extensively studied
\citep{Pastor-SatorrasVespignani:2000a,Pastor-SatorrasVespignani:2000e,HughesPaczuski:2002,PanETAL:2005b,JoHa:2008}. 
They have been characterised in detail by
\citet{PaczuskiBassler:2000} and related to the deterministic directed
models by \citet{Bunzarova:2010}. Exponents generally follow
$D=3/2+d_{\perp}/4$, which can be interpreted as the diffusive
exploration of a random environment. Again, correlations are suppressed
as sites are never re-visited in the same avalanche. As in the
deterministic case, $z=1$ and $D(2-\tau)=1$ and $D(\tau-1)=z(\alpha-1)$
result in $D=\alpha$. In $d=1+1$ exponents are $\tau=10/7$, $D=7/4$,
$\alpha=7/4$ and $z=1$.

\newcommand{\BFM}[1]{{\bf #1}}

\subsubsection{The Forest Fire Model}
The Forest Fire Model has an interesting, slightly convoluted history.
Two distinct versions exist, which share the crucial feature that
the bulk dynamics is not conservative. In the original version 
introduced by \citet{BakChenTang:1990} sites $i$, most frequently
organised in a
(two-dimensional) square lattice with periodic boundary conditions, can be in one of three states
$\sigma_i\in\{T,F,A\}$, corresponding to occupation by a {\BFM T}ree, by
{\BFM F}ire or by {\BFM A}sh. 
As time $t$ advances in discrete steps,
the state changes cyclically under certain
conditions: A {\BFM T}ree turns into {\BFM F}ire at time $t+1$ if a nearest
neighbouring site was on {\BFM F}ire at time $t$. In turn, a {\BFM F}ire
at time $t$ becomes {\BFM A}sh in time $t+1$, and a site covered in {\bf
A}sh at time $t$ might become occupied by a {\BFM T}ree at time $t+1$ due
to a repeated Bernoulli
trial with (small) probability $p$. 
\renewcommand{\BFM}[1]{\lowercase{#1}}
Starting from a lattice covered in
{\BFM T}rees, a single site is set on {\BFM F}ire and the system evolves
under the rules described. The key observable is the number of sites on
{\BFM F}ire as a function of time.

\begin{description}
\item[{\bf Initialisation:}] All (many) sites $i$ contain a {\BFM T}ree
(otherwise {\BFM A}sh),
$\sigma_i=T$, and (at least) one site is on {\BFM F}ire, $\sigma_i=F$.
\item[{\bf Driving:}] With (small) probability $p$, a site $i$ containing {\BFM A}sh at the beginning of
time step $t$ contains a {\BFM T}ree, $\sigma_i=A\to T$ at time
$t+1$.
\item[{\bf Toppling:}] A site $i$ that contains a {\BFM T}ree at begining
of time step $t$ and has at least one nearest neighbour on {\BFM F}ire,
turns into {\BFM F}ire as well, $\sigma_i=T\to F$.
Simultaneously, a site on
{\BFM F}ire at $t$ turns into {\BFM A}sh, $\sigma_i=F\to A$.
\item[{\bf Dissipation:}] {\BFM T}rees grow slowly in Bernoulli trials
and are removed in the ``toppling''. Their number is not conserved under
any of the updating.
\item[{\bf Time progression:}] Time progresses by one unit per parallel
update.
\end{description}

\begin{figure}[t]
\begin{center}
\includegraphics[width=0.5\linewidth]{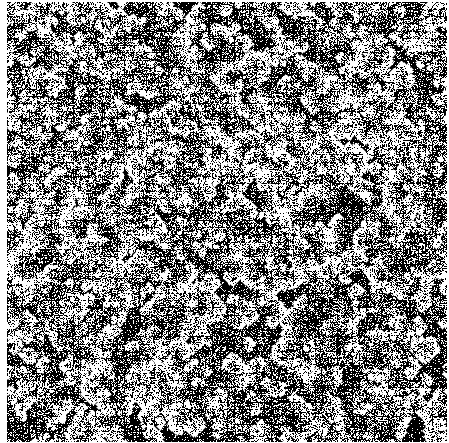}
\end{center}
\caption{
\flabel{ffm}
Realisation of the original Forest Fire Model by
\citet{BakChenTang:1990}. 
{\bf A}sh  is marked by a white site, {\bf T}rees are black and {\bf F}ires grey.
}
\end{figure}

The original \abbrevdef{Forest Fire Model}{FFM} just described
possesses an \keyword{absorbing state} from which it cannot recover within the
rules given. If the {\BFM F}ire stops spreading because the
last site on {\BFM F}ire is surrounded by {\BFM A}sh, the only transition
that can and will take place eventually occupies every site by a {\BFM T}ree.
\citet{BakChenTang:1990} originally suggested that occasional
re-lightning might be necessary --- in fact, if $p$ is large enough, on
sufficiently large lattices, there will always be {\BFM T}ree to burn
available. This, however, points to a fundamental shortcoming, as
quantified by \citet{GrassbergerKantz:1991}, namely that the lengthscale
of the relevant features of the FFM are determined by $p$. Typically, at
small $p$, some large spiral(s) of {\BFM F}ire keeps sweeping across the
lattice. If $p$ is chosen too small, the spatial extent of the spiral
becomes too large compared to the size of the lattice and the {\BFM F}ire
eventually goes out. However, if a control parameter determines the
characteristic length scale of the phenomenon, it cannot be \latin{bona
fide} SOC \citep[e.g.][]{BonachelaMunoz:2009}.
\Fref{ffm} shows an example of the structures, most noticeable
the fire fronts, developing.

The name ``Forest Fire Model'' should be taken as a witty aide-memoire.
\citet{BakChenTang:1990} designed the model to understand scale free
dissipation with uniform driving as observed in turbulent flow. The
model
should therefore be considered much more as a model of turbulence that
happened to look like fires spreading in a forest. In the present
model, perpetual fires spread across {\BFM T}rees as they re-grow, which is a
rather unrealistic picture; most fires in real forests are shaped by
fire brigades, geographical and geological features and other
environmental characteristics, as well as policies. Nevertheless, the
original FFM as well as the version by \citet{DrosselSchwabl:1992a},
attracted significant attention as an actual model of forest fires, as
well as other natural and sociological phenomena \citep{Turcotte:1999}.

There are two distinguishing features that set the FFM apart from many other
SOC models. Firstly, the separation of time scales is incomplete, because
driving the system by supplying new {\BFM T}rees is a process running in
parallel to the burning as {\BFM F}ire spreads. 
Although the time scale of {\BFM T}ree growth, parameterised by
$p$, can in principle be made arbitrarily slow, 
the {\BFM F}ire has to be constantly fed by new trees and cannot
be allowed to go out, because there is no explicit re-lighting. In other
words, the {\BFM T}ree growth rates that still sustain fire are bounded
from below.
As a result, there are no
distinct avalanches, as found in the BTW and the Manna Models.

More importantly, however, the FFM is different from other models
because it is non-conservative at a fundamental level. No quantity is
being transported to the boundaries and the local degree of freedom
changes without any conservation.\footnote{It is difficult to make the
statement about non-conservation strict. After all, the state of each
site is meant to change and allowing for that, it is always possible
to trace the appearance and the disappearance of something back to some
influxes and outfluxes. Here is an attempt in the present case: While
the increase in the number of {\BFM T}rees can be thought of as being
due to a corresponding influx, they can disappear with an enormous rate
by spreading {\BFM F}ire without explicit outflux \emph{on that
timescale}.} At the time of the introduction of the FFM, it challenged
\posscitet{HwaKardar:1989a} suggested mechanism of SOC that relied on a
conservation law to explain the absence of a field-theoretic mass in the
propagator. 

Other dissipative models, like the SOC version of the ``Game of Life''
\citep{BakChenCreutz:1989}, the OFC model discussed in the next section
\citep{OlamiFederChristensen:1992} and the Bak-Sneppen Model
\citep{BakSneppen:1993} chipped away from the conservation argument put
forward by \citet{HwaKardar:1989a,HwaKardar:1992}, \citet{GrinsteinLeeSachdev:1990} and
\citet{SocolarGrinsteinJayaprakash:1993}. The latter seem to have been
caught by surprise by the advent of a variant of the FFM by
\citet{DrosselSchwabl:1992a} discussed in the following.

The \abbrevdef{Drossel-Schwabl Forest Fire Model}{DS-FFM}, as it is now normally referred
to, was originally introduced by \citet{Henley:1989}. It changes the
original Forest Fire Model in two very important points: Firstly,
the separation of time scales between burning and growing is completed,
so that patches of (nearest neighbouring) {\BFM T}rees are burned down
instantly compared to all other processes. Because fires therefore burn
down completely before anything else can happen, fires are set,
secondly, explicitly by random, independent uniform lightning. The
key-observables of the DS-FFM are the geometrical features of the
clusters burned down, such as the number of occupied sites (the mass)
and the radius of gyration.

While {\BFM T}rees grow with rate $p$ on every empty site (\ie one containing
{\BFM A}sh), lightning strikes with much lower rate $f$ on every site.
If it contains a 
{\BFM T}ree, the fire eradicates the entire cluster of
{\BFM T}rees connected to it by nearest neighbour interactions. In
summary:
\begin{description}
\item[{\bf Initialisation:}] All sites $i$ contain {\BFM A}sh,
$\sigma_i=A$.
\item[{\bf Driving:}] With (small) probability $p$, a site $i$ containing {\BFM A}sh at the beginning of
time step $t$ contains a {\BFM T}ree, $\sigma_i=A\to T$ at time
$t+1$.
\item[{\bf Toppling:}] With probability $f\ll p$, a site containing a
{\BFM T}ree at the beginning of time step $t$ and the entire cluster of
{\BFM T}rees connected to it by nearest neighbour interactions is changed
to {\BFM A}sh, $\sigma_i=T\to A$.
\item[{\bf Dissipation:}] {\BFM T}rees grow slowly in Bernoulli trials
and are removed in the ``toppling''. They are not conserved in any of
the updates.
\item[{\bf Time progression:}] Time progresses by one unit per parallel
update, toppling is instantaneous relative to growing {\BFM T}rees. 
\end{description}

As a result
entire patches of forest disappear at a time, which are re-forested
with the same Poissonian density $p$. This process results in a patchy structure
with individual islands having roughly homogeneous tree-density, \Fref{ds_ffm}. 

\begin{figure}[t]
\begin{center}
\includegraphics[width=0.4\linewidth]{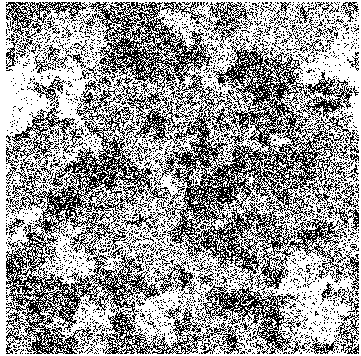}
\end{center}
\caption{
\flabel{ds_ffm}
Realisation of the Drossel-Schwabl Forest Fire Model
\citep{DrosselSchwabl:1992a}.
{\bf A}sh is marked by a white site, {\bf T}rees are black.
}
\end{figure}

In a change of perspective, the processes parameterised by $p$ and $f$
are {\BFM T}ree growth attempts and lightning attempts which fail if the site
is already occupied by a {\BFM T}ree or does not contain one,
respectively. The original definition by \citet{DrosselSchwabl:1992a}
still used discrete time, so that both $p$ and $f$ were probabilities,
rather than Poissonian rates, which can be recovered by rescaling $p$
and $f$ simultaneously. However, it is common
\citep[e.g.][]{ClarDrosselSchwabl:1996} to rescale time so that $p=1$
(enforced growth on randomly picked empty sites) and to attempt $p/f$
times to grow a {\BFM T}ree before
attempting to set one alight. In order to see scale-free cluster size
distributions, a \keyword{second separation of timescales} is needed,
whereby the ratio $p/f$ diverges.

Many of the properties of the DS-FFM are percolation-like. If it were not
for the correlations in the tree-density, which develop because of
``synchronous, patchy re-forestation'', i.e. if the tree-density was
homogeneous, then the DS-FFM would be a form of percolation. In
particular, the
cluster size distribution (of the patches removed and the totality of all patches
present) was given by that of (well-known) static percolation.

The DS-FFM does not suffer from the same short-coming as the original
FFM of having a well-understood typical (spiral) structure, whose size
is determined by the single control parameter $p$, yet it still has one
control parameter which needs to be finely tuned in accordance with the
system size. This parameter is $p/f$ --- if it is too large, then the
lattice will be densely filled with {\BFM T}rees before lightning strikes and
removes almost all of them, leaving behind essentially a clean sheet
with a few remaining (small) islands of {\BFM T}rees. If $p/f$ is too small,
then no dense forest ever comes into existence and the cluster size
distribution has a cutoff not determined by the system size, but by that
parameter.

In extensive numerical studies
\citep{Grassberger:2002a,JensenPruessner:2002b,JensenPruessner:2004},
the system sizes were chosen big enough for each $p/f$ that finite size
effects were not visible, \ie for each $p/f$ convergence of the cluster
size distribution $\pdf{}{s;L}$ in the system size $L$ was achieved.
However, these studies revealed that the DS-FFM does not display simple
scaling in $s_c=s_c(p/f)$, \Eref{simple_scaling} (\Sref{simple_scaling}). While
$\pdf{}{s}/s^{-\tau}$ converges in the thermodynamic limit (as it
should, trivially) for any $\tau$, there is no choice of $\tau$ so that
the remaining functional profile depends only on the ratio $s/s_c(p/f)$.
Instead, $\pdf{}{s}/s^{-\tau}$ depends explicitly on both $s$ and
$s_c(p/f)$, or, for that matter, $p/f$. The only feature that may
display some convergence \citep{JensenPruessner:2002b} is the bump in the
probability density function (PDF) towards large $s$. For some choice of $\tau$, there is a small
region, say $[s_c(p/f)/2, s_c(p/f)]$, where $\pdf{}{s}/s^{-\tau}$ traces out
a very similar graph, as if the lower cutoff $s_0$ itself was a
divergent multiple of the upper cutoff.\footnote{If $s_c(p/f)$ marks
roughly the maximum of the bump, the PDF drops off beyond it so
quickly, that next to nothing is known of $\pdf{}{s}$ beyond $s_c$. In
principle, however, if there is approximate coincidence on $[s_c(p/f)/2,
s_c(p/f)]$, there should also be approximate coincidence on
$[s_c(p/f)/2, \infty)$.}

One may hope that finite size scaling can be recovered, taking the limit
of large $p/f$ and considering $\pdf{}{s}/s^{-\tau}$ as a function of
$L$. However, it is clear that the PDF trivialises in this limit, 
\begin{equation}
\lim_{p/f\to\infty} \pdf{}{s;p/f,L} = s^{-1}
\delta\left(\frac{s}{L^d}\right)
\end{equation}
as the lattice is completely covered in {\BFM T}rees before they all get
completely removed in a singly lightning.

Interestingly, the lack of scaling in finite $s_c(p/f)$ is not visible
in the scaling of the moments $\ave{s^n}$ because they are sensitive to
large event sizes (at any fixed $n>\tau-1$), rather than the smaller
ones around the lower cutoff, whose divergence violates simple scaling.

As in the BTW Model, exponents reported for the DS-FFM (if they are
reported at all) display a fairly wide spread. In two dimensions, they
are $\tau$ from $1$ \citep{DrosselSchwabl:1992a} to $1.48$
\citep{PatzlaffTrimper:1994} and $D$ from $1$ \citep{DrosselSchwabl:1992a}
to $1.17(2)$ \citep{Henley:1993,HoneckerPeschel:1997}.

\subsubsection{The OFC Model}
To this day, the Olami-Feder-Christensen Model \citep[OFC
Model][]{OlamiFederChristensen:1992} is one of the most popular and spectacular models
of SOC. It is a simplified version of the Burridge-Knopoff Model
\citep{BurridgeKnopoff:1967} of earthquakes,
it has a \emph{tunable} degree of non-conservation (including a
conservative limit)  with a clear physical
meaning, it has been extensively analysed, both in time and space, for the effect of
different boundary conditions \citep{MiddletonTang:1995}, and its one-dimensional variant
\citep{deSousaVieira:1992} has been linked to the Manna universality
class \citep{PaczuskiBoettcher:1996,ChiancaSaMartinsdeOliveira:2009}.
After the definition of the model, the discussion below focuses on the
model's r{\^o}le in earthquake
modelling and the attention it received for the spatio-temporal patterns
it develops.

The OFC Model is at home on a two-dimensional square lattice. As in the
models above, each site $i$ has a local degree of freedom
$z_i\in\Rset$ (called the local ``pulling force''),
which is, in contrast to the models above, however, real-valued. As in
the BTW Model, there are two clearly distinct stages of external driving
and internal relaxation. During the driving \emph{all} sites in the
system receive the same amount of force (sometimes referred to as
``continuous'' or better ``uniform'' drive) until one site exceeds the
threshold $z^c=1$, which triggers a relaxation during which no further
external driving is applied. In a relaxation or
toppling, a site re-distributes a fraction of \emph{all} pulling force evenly among its
nearest neighbours which may in turn exceed the threshold. 
The force $z_i$ at the toppling site $i$ is set to $0$ and the amount arriving at
each neighbour is $\alpha z_i$, where $\alpha$ is the \keyword{level of
conservation}. At coordination number $q$, a level conservation less
than $1/q$ means that the bulk dynamics is dissipative. Boundary sites
lose force $\alpha z_i$ (at corners multiples thereof) to the outside.
Because the
force re-distributed depends on the amount of pulling force present at the site at the time of the
re-distribution, the order of updates matters greately, \ie the OFC Model
is not Abelian. If $\alpha<1/q$ periodic boundary conditions can be
applied without losing the possibility of a stationary state, yet
normally the boundaries are open. The OFC Model is normally initialised
by assigning random, independent forces from a uniform distribution.

Sites to topple are identified at the beginning of a timestep and only
those have been relaxed by the end of it (parallel updates). 
Unless more than one site exceeds the threshold (degenerate maximum) at the beginning of an
avalanche, toppling sites therefore reside on either of the two next
nearest neighbour sublattices of a square lattice.

Again, a separation of time scales is applied, where the relaxation
becomes infinitely fast compared to an infinitesimally slow drive. In an actual
implementation, however, the driving is applied instantaneously and the relaxation
takes up most (computational time): The driving can be completed in a
single step by keeping track of the site, say $i^*$ with the largest pulling force
acting on it. The amount of force added throughout the system is thus
simply $z^c-z_{i^*}$, triggering the next avalanche.

Because sweeping the lattice in search of the maximum is
computationally very costly,\footnote{Not only is the very searching
\emph{across all sites} costly, most of the memory occupied by the lattice
will not reside in a cache line (as for example most ``local'' data) 
and thus has to be fetched through a
comparatively slow bus.}
the main computational task in the OFC
Model is to keep track of the site exposed to the maximum pulling force.
This is a classic computational problem
\citep{CormenLeisersonRivest:1996}, which also is occurs in other models,
such as the Bak-Sneppen Model \citep{BakSneppen:1993}. The traditional solution is to
organise data in a tree-like structure and devise methods that allow
fast updating and determination of the maximum. However, in the OFC
Model updating as site's force is much more frequent than determination of the maximum
and thus a fast algorithm focuses on the optimisation of the former
at the expense of the latter, \ie a slightly slower procedure to determine the maximum.

\citet{Grassberger:1994} pointed out a number of improvements to a
na{\"i}ve, direct implementation of the OFC Model. Firstly, instead of
driving the system uniformly, thereby having to sweep the lattice to
increase the force on every site by $z^c-z_{i^*}$, the threshold $z^c$
is to be lowered; the amount of force re-distributed at toppling is
obviously to be adjusted according to the new offset. The second major
improvement \citet{Grassberger:1994} suggested was the organisation of
forces in ``boxes'' (sometimes referred as \keyword{Grassberger's
box-technique}), which splits the range of forces present in the system
in small enough intervals that the search for the maximum force succeeds
very quickly, yet keeps the computational effort to a minimum when re-assigning a box
after an update. 
Other improvements suggested was maintaining a stack
(\Sref{stack_discussion}) of active sites, and the use of a scheme to
determine neighbouring sites suitable to the programming language at
hand.

The adjustment of $z^c$ outlined above has some rather unexpected effects
depending on the numerical precision (\Sref{numerical_precision}) used
in the simulation \citep{Pruessner:2012:Book}. As pointed out by
\citet{Drossel:2002}, the OFC Model is extremely sensitive to a change
of precision; a lower precision seems to enhance or favour
phase-locking, discussed in the following.

Most of the studies of SOC models focuses on large-scale statistical
features, large both in time and space. The analysis of the OFC Model by
\citet{SocolarGrinsteinJayaprakash:1993}
\citet{MiddletonTang:1995} and \citet{Grassberger:1995} therefore
ventured into uncharted territory as they
studied the evolution towards stationarity in the OFC Model on a
microscopic
scale, analysing the patchy structure of the forces on the lattice.

Firstly, periodic boundary conditions inevitably
lead to periodic behaviour in time. Below $\alpha\approx0.18$ in a
two-dimensional square lattice, (almost) every avalanche has size unity.
In that extreme case, the period is strictly $1-q\alpha$, because
discounting the external drive, this
is the amount of force lost from every site after every site has toppled
exactly once, as the system goes through one full period. 

The periodicity is broken once open boundaries are introduced. Sites
at the edge of the lattice have fewer neighbours that charge them, so if
every site in the system topples precisely once, the force acting on a
boundary site is expected to be lower. While open boundaries indeed
break temporal periodicity, they form, at the same time, seeds for
(partially) synchronised patches, which seem to grow from the outside
towards the inside, increasing in size. \citet{MiddletonTang:1995} introduced the term
\keyword{marginal (phase) locking} to describe this phenomenon. 

The temporal periodicity might similarly be broken by introducing inhomogeneities
or disorder, effective even at very low levels
\citep{Grassberger:1994,Ceva:1995,Ceva:1998,TorvundFroyland:1995,MiddletonTang:1995,Mousseau:1996}.
That a spatial inhomogeneity helps forming synchronised patches
in space can also be attributed to marginal phase locking. 

Because the OFC Model is so sensitive to even the smallest amount of
disorder and inhomogeneity, its statistics is often taken from very big
samples with extremely long transients. Many authors also average over
the initial state. \citet{Drossel:2002} suggested that despite these
precautions, some of the statistical behaviour allegedly displayed
by the OFC Model might rather be caused by numerical ``noise'', also a
form of inhomogeneity or disorder entering into a simulation. In
practise, it is difficult to discriminate genuine OFC behaviour from
numerical shortcomings and one may wonder whether some of these
shortcomings are not also present in the natural phenomenon the OFC
Model is based on.

That SOC may be applicable in seismology had been suggested by
\citet[][also
\citealp{BakTang:1989a,SornetteSornette:1989,ItoMatsuzaki:1990}]{BakTangWiesenfeld:1989} at
a very early stage. The breakthrough came with the OFC Model, which is
based on the Burridge-Knopoff Model of earthquakes (or rather fracturing rocks). The latter is
more difficult to handle numerically, with a ``proper'' equation of
motion taking care of the loading due to spring-like interaction much
more carefully.
The OFC Model, on the other hand, is much easier to update, almost like
a  cellular automaton.\footnote{Strictly, the OFC Model generally is not a cellular
automaton, because the local states $z_i$ are continuous.} The context
of SOC provided an explanatory framework of the scale-free occurrence of
earthquakes as described by the \keyword{Gutenberg-Richter} law
\citep{GutenbergRichter:1954,OlamiFederChristensen:1992}. Even though
exponents both in the real-world as well as in the OFC Model seem to
lack universality, certain scaling concepts, motivated by studies in
SOC, have been applied successfully to earthquake catalogues
\citep{BakETAL:2002a}.

It is fair to
say that the OFC Model, to this day, is widely disputed as a \latin{bona
fide}
model of earthquakes. Its introduction has divided the seismology
community, possibly because of the apparent disregard 
of their achievements
by the proponents
of SOC \citep{BakTang:1989a}. One of the central claims made initially
is that earthquakes are unpredictable if they are ``caused'' by SOC,
which questions the very merit of seismology. However, given that SOC
is a framework for the understanding of natural phenomena on a long time
and length scale, providing a mechanism for the existence of long
temporal correlations, SOC indicates precisely the opposite of
unpredictability. This point is discussed controversially in the
literature to this day
\citep{Corral:2003,Corral:2004a,Corral:2004b,DavidsenPaczuski:2005,LindmanETAL:2005,CorralChristensen:2006,LindmanETAL:2006,WernerSornette:2007,DavidsenPaczuski:2007,SornetteWerner:2009}.
Older reviews \citep{Turcotte:1993,CarlsonLangerShaw:1994} help to
understand the historical development of the dispute.
\citet{Hergarten:2002} and more recently \citet{SornetteWerner:2009}
have put
some of the issues in perspective.

There is not a single set of exponents for the OFC Model, as they are
generally expected to vary with the level of conservation
\citep{ChristensenOlami:1992a}. Because authors generally do not agree on the
precise value of $\alpha$ to focus on, results are not easily comparable
across studies. Even in the conservative limit, $\alpha=1/q$, little
data is available, suggesting $\tau=1.22(5)--1.253$ and
$D=3.3(1)--3.01$ \citep{ChristensenOlami:1992a,ChristensenMoloney:2005}.

\section{Scaling and numerics}
\slabel{scaling}
As a rule of thumb, SOC models are \keyword{SDIDT} systems
\cite{Jensen:1998}: {\bf S}lowly {\bf D}riven {\bf I}nteraction {\bf
D}ominated {\bf T}hreshold systems. The driving implements a separation
of time scales and thresholds lead to highly non-linear interaction, which
results in avalanche-like dynamics, the statistics of which displays
scaling, a continuous symmetry. Ideally, the scaling behaviour
of an SOC model can be related to some underlying continuous phase
transition, which is triggered by the system self-organising to the
critical point.

The critical behaviour can be characterised by (supposedly) universal
critical exponents, the determination of which is the central theme of the
present section. At the time of the conception of SOC, critical
exponents were extracted directly from probability density function,
(PDFs),
often by fitting the data to a straight line in double-logarithmic plot.
Frequently, such scaling is referred to as ``power law behaviour''. Very much
to the detriment of the entire field, some authors restrict their
research to the question whether an observable displays the desired
behaviour, without attempting to determine its origin and without
considering the consequences of such behaviour. Power law behaviour
therefore has become, in some areas, a mere curiosity.

\subsection{Simple scaling}
\slabel{simple_scaling}
While studying power laws in PDFs can be instructive, there are far
superior methods to quantify scaling behaviour. In recent years, most
authors have focused on an analysis of the moments of the PDFs, as
traditionally done in the study of equilibrium statistical mechanics.
Not only is this approach more efficient, it also is more accurate and
mathematically better controlled. Moreover, it is concerned directly
with an observable (or rather, arithmetic means of its powers), rather than its
accumulated histogram.

Nevertheless the starting point of a scaling analysis in SOC, is the
\keyword{simple (finite size) scaling assumption},
\begin{equation}
\pdf{}{s} = a s^{-\tau} \GC(s/s_c) \text{ for } s\gg s_0 \ ,
\elabel{simple_scaling}
\end{equation}
where $\pdf{}{s}$ is the (normalised) probability density function of an
observable, $s$ in this case, $a$ is a (non-universal) \keyword{metric factor} present to restore
dimensional consistency and accounting for the (microscopic) details of
the model, $\tau$ is a \keyword{universal scaling} (or critical)
\keyword{exponent}, $\GC$ is a \keyword{universal scaling function}, $s_c$ is the
\keyword{upper cutoff} and $s_0$ the \keyword{lower cutoff}. If $s$ is the
avalanche size, then $\tau$ is known as the \keyword{avalanche size
exponent}, when $s$ is the duration, then $\tau$ is traditionally
replaced by $\alpha$ and called the \keyword{avalanche duration
exponent}.

The two cutoffs
signal the onset of new physics: Below $s_0$ some microscopic physics
prevails, often a lattice spacing or some other minimal length below
which discretisation effects take over. Above $s_c$ some large finite
length scale becomes visible, which in SOC is normally controlled by the
size of the lattice, so that \Eref{simple_scaling} is referred to as
\emph{finite size} scaling. In traditional critical phenomena, $s_c$ is
controlled by the correlation length, beyond which distant parts of the
system can be thought of as being independent, suggesting the validity
of the central limit theorem.

Strictly, SOC models should always tune themselves to a critical point,
so that the algebraic, critical behaviour is cut off only by the system
size. All scaling in SOC therefore is finite size scaling. There are a
handful of established SOC models, which violate that strict rule, however, such as
the Drossel-Schwabl Forest Fire Model \cite{DrosselSchwabl:1992a}, where
an additional parameter has to be tuned simultaneously with the system
size. 

The physical origin of the scales contained in the metric factor $a$ and
the lower cutoff $s_0$ often is the same, yet even with these length scales present, $\pdf{}{s}$
has an arbitrarily wide region where it displays a power-law dependence
on $s$ and whose width is controlled by $s_c$; if $s_0\ll s\ll
s_c$, then $\pdf{}{s} = a s^{-\tau+\alpha} s_c^{-\alpha} \GCtilde_0$, provided 
\begin{equation}
\lim_{x\to0} x^{-\alpha} \GC(x) = \GCtilde_0 \ .
\end{equation}
Typically, however, $\alpha=0$ so that the intermediate region of
$\pdf{}{s}$ displays a power law dependence with exponent $\tau$, which
can in principle be extracted as the negative slope of $\pdf{}{s}$ in a double logarithmic plot.
However, because it is \latin{a priori} unclear whether the scaling
function $\GC(s/s_c)$ can
be approximated sufficiently well by a constant $\GC_0$, ``measuring''
the exponent $\tau$ by fitting the intermediate region of a double
logarithmic plot to a straight line (sometimes referred to as the
\keyword{apparent exponent}) is very unreliable. If the scaling function displays a
power law dependence on the argument, $\alpha\ne0$, the effective exponent in the
intermediate region is $\tau-\alpha$. One can show that $\alpha$ is
non-negative, $\alpha\ge0$, and $\tau=1$ if $\alpha>0$
\citep{ChristensenETAL:2008}.

\begin{figure}
\begin{center}
\includegraphics[width=0.7\linewidth]{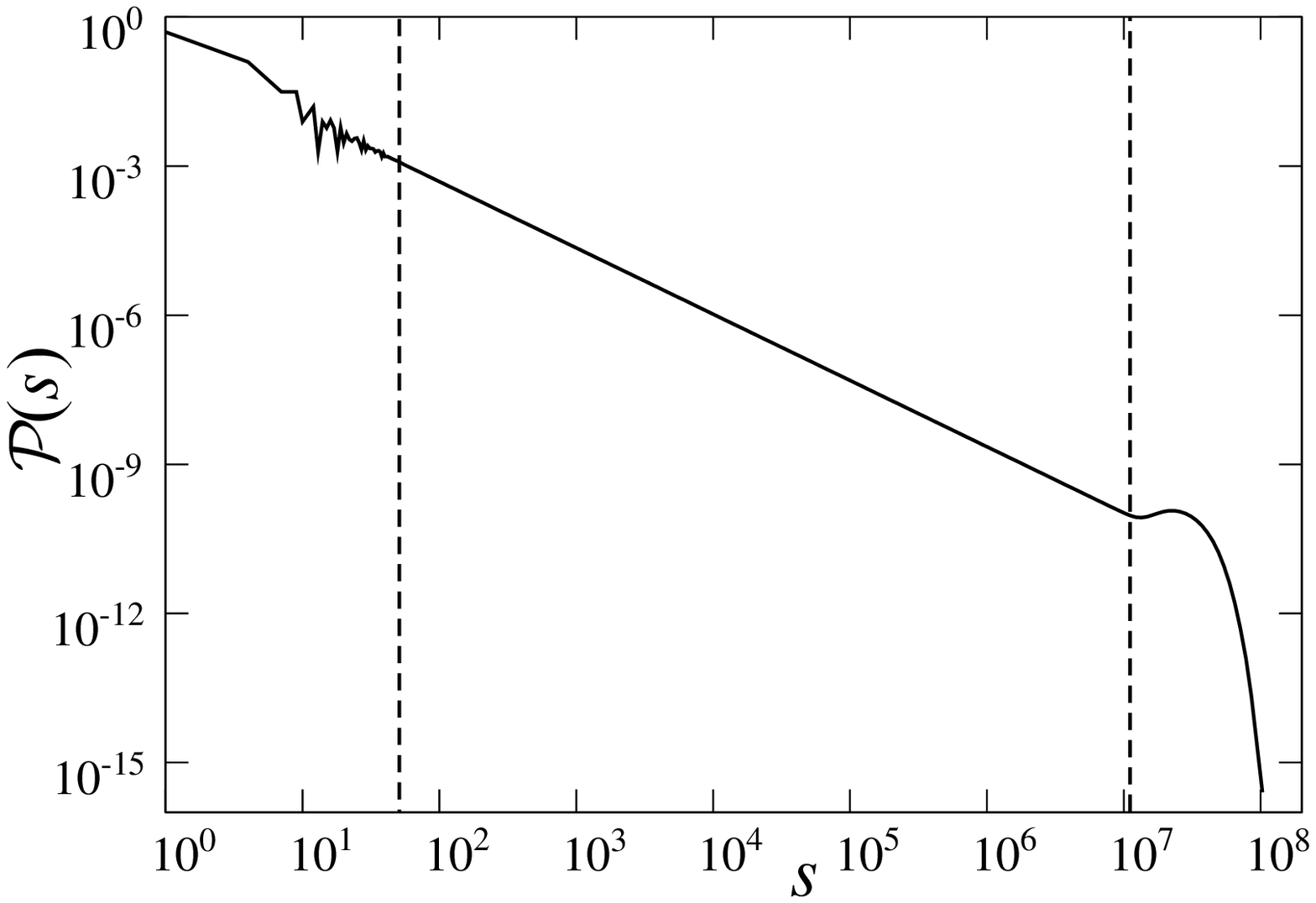}
\end{center}
\caption{\flabel{rw_pdf_bare_labelled}Example of a double logarithmic plot of the PDF of the
avalanche size in an
SOC model \citep[Data from][]{Pruessner:2012:Book}.}
\end{figure}

\Fref{rw_pdf_bare_labelled} shows a typical double-logarithmic plot of
the PDF in an SOC model. The power law region is marked by two dashed
lines. The lower cutoff is at around $s_0=50$ and the features below
that value are expected to be essentially reproduced by that model
irrespective of its upper cutoff. The spiky structure visible in that
region is not noise and may, to some extent, be accessible analytically,
similar to the lattice animals known in percolation
\citep{StaufferAharony:1994}.
The power law region between the two dashed lines can be widened
arbitrarily far by increasing the upper
cutoff $s_c$. Running the same model with increasing $s_c$ will
reproduce this almost straight region beyond which the bump in the data indicates the
onset of the upper cutoff. 

The upper cutoff in SOC models supposedly depends only on the system
size and does so in a power-law fashion itself,
\begin{equation}
s_c(L) = b L^{D}
\elabel{supcut_scaling}
\end{equation}
where $b$ is another metric factor and $D$ is the \keyword{avalanche
dimension}. The exponent describing the same behaviour for the upper
cutoff of the avalanche duration is the \keyword{dynamical exponent}
$z$. The four exponents $\tau$, $D$, $\alpha$ and $z$ are those most
frequently quoted as the result of a numerical study of an SOC model.

The simple scaling ansatz \Eref{simple_scaling} as well the scaling of
the upper cutoff, \Eref{supcut_scaling}, both describe \emph{asymptotic}
behaviour in large $s_c$ and $L$ respectively. When determining
exponents in computer simulations of SOC models, corrections have
to be taken into account in a systematic manner. While subleading terms
are difficult to add to the simple scaling ansatz \Eref{simple_scaling},
this is routinely done in the case of the upper cutoff,
\begin{equation}
s_c(L) = b L^{D} (1 + c_1 L^{-\omega_1} + c_2 L^{-\omega_2}\ldots)
\elabel{full_supcut_scaling}
\end{equation}
Corrections of this form are referred to as \keyword{corrections to
scaling} \citep{Wegner:1972} or confluent singularities. These corrections
are discussed further in the context of moment analysis,
\Sref{moment_analysis}.

\begin{figure}
\begin{center}
\includegraphics[width=0.7\linewidth]{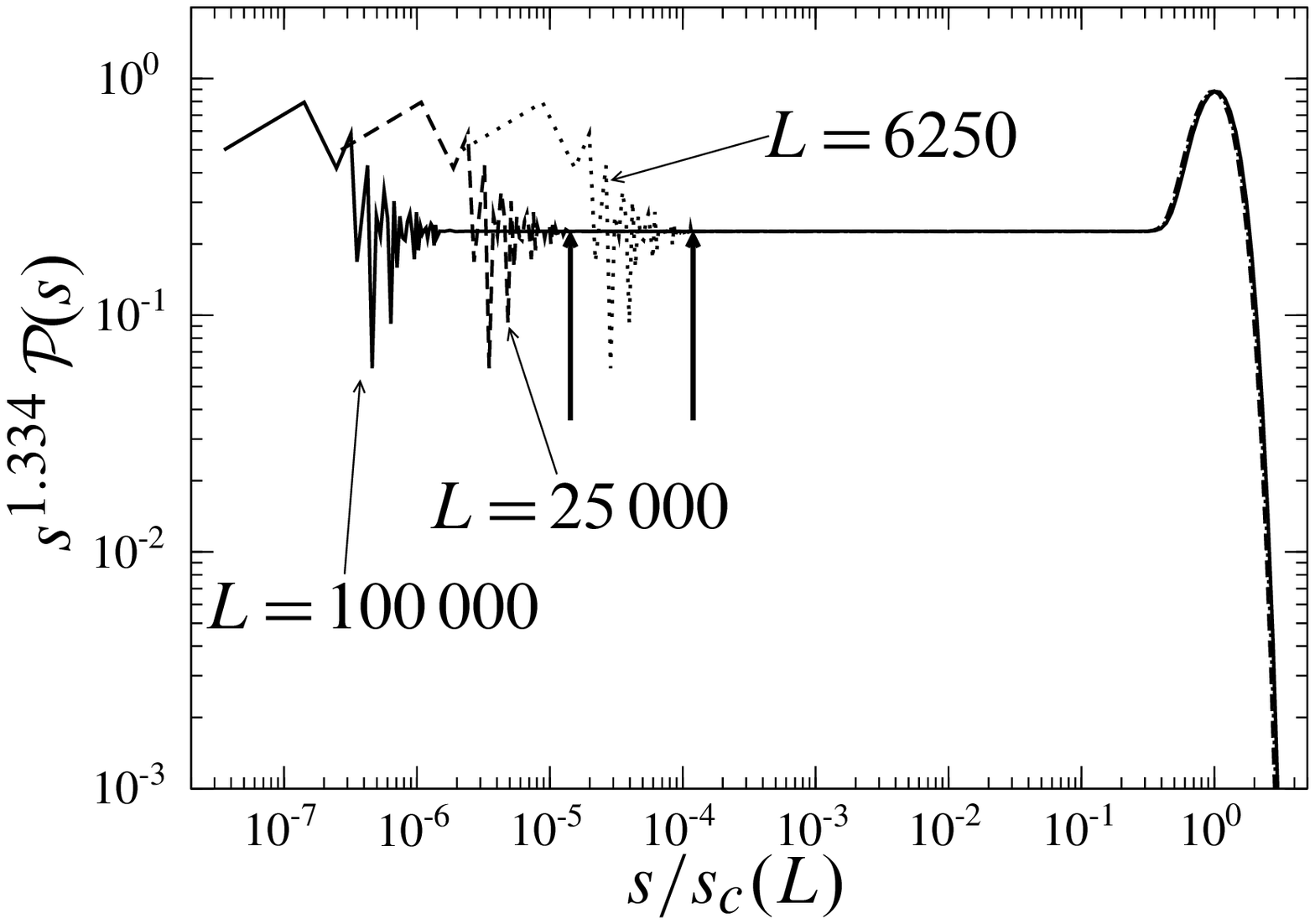}
\end{center}
\caption{\flabel{data_collapse}
Data collapse of three different data sets similar to the data shown in
\Fref{rw_pdf_bare_labelled}. The upper cutoff $s_c$ is solely controlled
by the system size $L$ \citep[Data from][]{Pruessner:2012:Book}.
}
\end{figure}

Although some very successful methods of analysis exist
\citep{ClausetShaliziNewman:2009},
\Eref{simple_scaling} does not lend itself naturally to a systematic
quantitative analysis for fixed $s_c$. Often, a \keyword{data collapse} is performed in
order to demonstrate the consistency of the data with simple scaling.
According to \Eref{simple_scaling} the PDF $\pdf{}{s}$ for different
cutoffs $s_c$ produces the same graph by suitable rescaling, in
particular by plotting $\pdf{}{s} s^\tau$ against $x=s/s_c$, which gives
$\GC(x)$. Deviations are expected for small values of $s/s_c$, namely
for $s$ around $s_0$, where \Eref{simple_scaling} does not apply.
\Fref{data_collapse} shows an example of such a collapse using the same
model as in \Fref{rw_pdf_bare_labelled}.

Provided $\lim_{x\to0} \GC(x)=\GC_0\ne0$, the region where $\pdf{}{s}$
displays (almost) a power law translates into a horizontal, (nearly)
constant section in the rescaled plot. The graph terminates in a
characteristic \keyword{bump}, where the probability density of some larger event
sizes
exceeds that of some large, but smaller ones. This counter-intuitive
feature is normally interpreted as being caused by system spanning
events which were terminated prematurely by the boundaries of the
system. Had the system been larger, the events would have developed
further.
In the PDF of a larger system thus make
up the regular, straight power law region, where the smaller system's
PDF displays a bump. Even when the total
probability contained in the bump is finite but very small, it is enough
to account for all events contained beyond it in the power law region of
an infinite system.

A data collapse is not unique, as
plotting  $\pdf{}{s} s^\tau f(s/s_c)$ produces $\GC(x)f(x)$ for any
function $f(x)$. In the literature, $f(x)$ is often chosen as
$f(x)=x^{-\tau}$ so that $\pdf{}{s} s^\tau f(s/s_c)=\pdf{}{s} s_c^\tau$.
Plotting that data has the fundamental disadvantage that $\pdf{}{s}
s_c^\tau$ usually spans many orders of magnitude more across the
ordinate compared to $\pdf{}{s} s^\tau$, so that details in the terminal
bump are less well resolved.

\subsubsection{Binning}
\slabel{binning}
A clean, clear dataset like the one shown in \Fref{rw_pdf_bare_labelled}
is the result of \keyword{binning}. For numerical studies of SOC models
this is a necessary procedure in order to smoothen otherwise rather
rugged histograms. The reason for that ruggedness is the strong
dependence of the probability density on the event size, with very few
large events occurring. Because of the power law relationship between
event size and frequency, their total numbers decrease even on a
logarithmic scale. As a result, statistical noise visibly takes over,
often clearly before the onset of the terminal bump. Statistics for
large event size is sparse and often little more than a muddle of
seemingly unrelated data points is visible in the raw data for large
events.

The noise can be reduced by averaging the data for
increasingly large event sizes over increasingly large ``bins'', hence
the name binning. This is normally done in \keyword{post-processing} of the raw data
produced in a numerical simulation, by summing over all events
within a bin and dividing by its size.
In principle, the bin sizes could be chosen to fit the
data; if the bin ranges are $[b_i,b_{i+1})$, then a pure power law
$\pdf{}{s}=a s^{-\tau}$ would deposit 
\begin{equation}
\int_{b_i}^{b_{i+1}} \dint{s} a s^{-\tau} = \frac{a}{\tau-1} \left(
b_{i}^{1-\tau} - b_{i+1}^{1-\tau} \right) 
\elabel{bin_filling}
\end{equation}
events in each bin $i$. This number can be made constant by choosing
$b_i=(B_0 - B_1 i)^{1/(1-\tau)}$. Similarly, one might chose the bin
boundaries $b_i$ ``on the fly'', \ie successively increase the bin size
until roughly a given number of entries have been collected. While those
two choices lead to uniformly low statistical errors (assuming constant
correlations), they both suffer
from significant shortcomings. Firstly, the exponent $\tau$ to be
estimated from the data should not enter into the very preparation of
the data that is meant to produce the estimate. This problem is mitigated by the
fact that $\tau$ may be determined through a separate, independent
procedure. Secondly and more importantly, both procedures will lead to
an increasingly wide spacing of data points, which becomes unacceptable
towards large event sizes, because the abscissa will no longer be
defined well enough --- if $b_{i+1}$ and $b_i$ are orders of magnitude
apart, which $s$ does \Eref{bin_filling} estimate. Last but not least, to make PDFs of different system sizes
comparable, the same $b_i$ should be used for all datasets.

The widely accepted method of choice is \keyword{exponential binning} (sometimes
also referred to as \keyword{logarithmic binning}), where $b_i=B_0
\exp(\beta i)$. Such bins are equally spaced on the abscissa of a double
logarithmic plot. Because the width of exponential bins is
proportional\footnote{For integer valued bin boundaries, strictly, this
holds only approximately.} to their limits, \Eref{bin_filling}, sparse
data can cause a surprising artefact, whereby single events spuriously
produce a probability density which decays inversely with the event
size, $\pdf{}{s}\propto s^{-1}$, suggesting an exponent of $\tau=1$.
A typical problem with exponential bins occurs at the small end of the
range when used for integer valued event sizes, because in that case the
$b_{i+1}-b_i$ should not be less than $1$. It is then difficult to
control the number of bins and thus the resolution effectively, because
decreasing $\beta$ increases the number of minimally sized bins and has
highly non-linear knock-on effects on all bin boundaries. The problem is
obviously much less relevant for non-integer event sizes, such as the
avalanche duration. However, it is rather confusing to use non-integer
bin boundaries for integer valued event sizes, because bins may remain
empty and the effective bin size cannot be derived from $b_{i+1}-b_i$.
For example a bin spanning $b_{i+1}-b_i=0.9$ may not contain a single
integer, whereas $b_{i+1}-b_i=1.1$ may contain two.

It is obviously advantageous to perform as much as possible of the data
manipulation as post-processing of raw simulation data. Efficiency and
memory limitations, however, normally require a certain level of binning
at the simulation stage. When event sizes and frequencies spread
over $10$ orders of magnitude a simple line of code\footnote{All explicit examples in this chapter are written in
C, which is the most widely used programming language for numerical
simulations, as long as they are not based on historic Fortran code.}
\begin{lstlisting}
histogram[size]++; /* one count for size in the histogram */
\end{lstlisting}
would require \code{histogram} to have a precision of more than $32$
bits. Normally such counters are implemented as integers, which would
need to be a \code{long long int} in the present case. The memory
required for $10^{10}$ of these $64$ bit numbers (about $75$ GB) exceeds
by far the memory typically available in computers in common use at the
time of writing this text (2012). Writing every event size in a list,
eventually to be stored in a file, is rarely an alternative, again because of
the enormous memory requirements and because of the significant amount
of computational time post-processing would take.

Consequently, some form of binning must take place at the time of the
simulation. In principle, any sophisticated binning method as used
during post-processing can be deployed within the simulation, yet the risk of coding errors
ruining the final result and the computational effort renders this
approach unfeasible. The established view that complicated floating
point operations such as \code{log} or \code{pow} are too expensive to
be used regularly in the course of a numerical simulation has
experienced some 
revision over the last decade or so, as techniques like hyperthreading
and out-of-order execution are commonly used even in the FPU.
Nevertheless, integer manipulation, often doable within a
single CPU cycle, remains computationally superior compared to floating
point manipulation. Even some of the rather archaic 
rules remain valid, such as multiplications being computationally more
efficient than divisions, as they can be performed within a short, fixed
number of cycles. Further details can be found in the appendix at the
end of the chapter.

\subsection{Moment analysis}
\slabel{moment_analysis}
By far the most powerful technique to extract universal features of an
SOC model is a moment analysis \citep{DeMenechStellaTebaldi:1998}. Traditionally, the numerical
investigation of critical phenomena has focused on moments much
more than on the underlying PDF, even when the former are often seen as
the ``result'' of the latter. Mathematically, no such primacy exists and
one can be derived from the other under rather general conditions
\citep[Carleman's theorem in]{Feller:1966}. In general one expects that
a finite system produces only finite event sizes, \ie that finite
systems have a sharp cutoff of the ``largest possible event size''.
While very physical, this rule finds its exception in residence times,
when particles get ``buried'' in a ``pile'' over long periods. In the
Oslo Model, some of these waiting time distributions seem to be
moderated by scaling functions that are themselves power laws and may
possess upper cutoffs exponential in the system size
\citep{DharPradhan:2004,PradhanDhar:2006,PradhanDhar:2007,Dhar:2006}.

Assuming, however, that all moments 
\begin{equation}
\ave{s^n}=\int_0^\infty \dint{s} s^n \pdf{}{s} 
\end{equation}
exist, \ie are finite, then for $n+1>\tau$
\begin{equation}
\ave{s^n} \simeq a s_c^{1+n-\tau} \gs{n}
\elabel{moment_fss}
\end{equation}
where $\simeq$ is used to indicate equivalence \emph{to leading order in
large $s_c$}. Moments with $n<\tau-1$ are not dominated by the scaling
in $s_c$, \ie they are convergent in large $s_c$. The (asymptotic) amplitudes $\gs{n}$
are defined as
\begin{equation}
\gs{n}=\int_0^\infty x^{n-\tau} \GC(x) 
\end{equation}
expected to be finite for all $n\ge0$. There is an unfortunate confusion
in the literature about the (spurious) consequences of $\ave{s^0}=1$
scaling like $s_c^{1-\tau} \gs{0}$. If $\tau>1$, then the leading order
of $\ave{s^0}$ is not given by \Eref{moment_fss}. 

The only scaling in SOC is finite size scaling, \ie the upper cutoff is
expected to diverge with the system size, \Eref{supcut_scaling}, so that
moments scale like
\begin{equation}
\ave{s^n} \simeq a b^{1+n-\tau} L^{D(1+n-\tau)} \gs{n} \ .
\end{equation}
Neither $a$ nor $b$ are universal and neither are the $\gs{n}$ unless
one fixes some features of $\GC(x)$ such as its normalisation and its
maximum. To extract universal characteristics of $\GC(x)$, moment ratios
can be taken for example
\begin{equation}
\frac{\ave{s^{n-1}} \ave{s^{n+1}}}{\ave{s^n}^2} = 
\frac{\gs{n-1} \gs{n+1}}{\gs{n}^2} + \text{corrections}
\end{equation}
or
\begin{equation}
\frac{\ave{s^n}\ave{s}^{n-2}}{\ave{s^2}^{n-1}} = 
\frac{\gs{1}^{n-2}}{\gs{2}^{n-1}} \gs{n} + \text{corrections} \ ,
\elabel{def_momrats}
\end{equation}
which is particularly convenient because of its very simple form when
fixing $\gs{1}=\gs{2}=1$ by choosing the metric factors $a$ and $b$
appropriately.

The most important result of a moment analysis, however, are the
universal exponents $D$ and $\tau$ and corresponding pairs for
avalanche duration ($z$ and $\alpha$ respectively), as well as the area
(normally $D_a$ and
$\tau_a$) \etc. This is done in a three step process.
Firstly, the SOC model is run with different systems sizes, typically
spaced by a factor $2$, or $2,5,10$. It can pay to use slightly
``incommensurate'' system sizes to identify systematic effects, for
example due to boundary effects being particularly pronounced in system
sizes that are powers of $2$. A typical simulation campaign
would encompass $10$ to $15$ system sizes, of which maybe only $6$ to
$10$, stretching over two to four orders of magnitude\footnote{One might
challenge the rule of thumb of the linear system size $L$ having to span
at least three orders of magnitude --- in higher dimensions, say $d=5$,
spanning three orders of magnitude in linear extent leads to $15$ orders
of magnitude in volume, which might be the more suitable parameter to
fit against.} 
will be used to produce estimates of exponents. The result of the
simulation are estimates for the moments of the relevant observables
together with their error (see below).

Secondly, the moments of the event sizes distribution, $\ave{s^n}$, are
fitted against a power law in $L$ (which is the parameter controlling
$s_c$) with corrections,
\begin{equation}
\ave{s^n}=A_0 L^{\mu_n} + A_1 L^{\mu_n-\omega_1} + \ldots
\elabel{fitting_function}
\end{equation}
with positive exponents $\omega_i$, known as confluent singularities; in
particular $\mu_n-\omega_1$ is sometimes referred to as a sub-dominant
exponent. The introduction of such \keyword{corrections to scaling} goes
back to \citet{Wegner:1972}, who applied it in equilibrium critical
phenomena. The Levenberg-Marquardt algorithm \citep{PressETAL:2007} is
probably the fitting routine most frequently employed for matching the
estimates (with their error bars) from the simulation to the fitting
function \Eref{fitting_function}. There are a
number of problems that can occur:
\begin{itemize}
\item Unless the result is purely qualitative, a good quality fit cannot
be achieved without good quality numerical data, that includes a solid
estimate of the numerical error, \ie the estimated standard deviation of the
estimate.
\item The very setup of fitting function \Eref{fitting_function}
(sometimes referred to as ``the model'')
can introduce a systematic error; after all it is only a hypothesis.
\item If $n>\tau-1$ is very small, corrections due to the presence of
the
lower cutoff ($s_0$, \Eref{simple_scaling}) can be very pronounced. 
\item The error stated for the fitted exponents alone can be misleading.
If \Eref{fitting_function} is very constraining, the error will be
low, but so will the goodness-of-fit.
\item Too many fitting parameters allow for a very good goodness of fit,
but also produce very large estimated statistical errors for the exponents.
\item Fitting against a function with many parameters often is
highly dependent on the initial guess.
In order to achieve good convergence and systematic, controlled results, it may pay off to fit the
data against \Eref{fitting_function} step-by-step, using the estimates
obtained in a fit with fewer corrections as initial guesses for a fit
with more corrections.
\item In most cases, there is little point in having as many parameters
as there are data points, as it often produces a seemingly perfect fit
(goodness-of-fit of unity), independent of the input data.
\item Extremely accurate data, \ie estimates for the moments with
very small error bars, may require a large number of correction terms.
\item It can be difficult to force the corrections $\omega_i$ to be
positive. It is not uncommon to fix them at certain reasonable values such as $\omega_i=i$
or $\omega_i=i/2$. Alternatively, they can be introduced differently,
writing them, for example, in the form $\omega_i=i+|\tilde{\omega}_i|$.
\item If finite size scaling applies, the relative statistical error for
any moment scales like $\ave{s^{2n}}/\ave{s^n}^2\propto L^{D(\tau-1)}$, assuming that
$\sigma^2(s^n)$ scales like $\ave{s^{2n}}$, which it certainly does for
$\tau>1$. At $\tau=1$ the scaling of 
$\sigma^2(s^n)$ may be slower than that of
$\ave{s^n}^2$. While $L^{D(\tau-1)}$ does not depend on $n$, the
amplitude of the moments does, leading normally to an increase of the relative
error with $n$.
\end{itemize}

In some models the first moment of the avalanche size displays
anti-correlations and thus faster numerical convergence as found in
a mutually independent sample \citep{WelinderPruessnerChristensen:2007}.
In many models, the average avalanche size $\ave{s}$ is known exactly,
in one dimension often including the confluent singularities
\citep{Pruessner_aves:2012}. These
exact results can provide a test for convergence in numerics and also
provide a \keyword{scaling relation}
\begin{equation}
D(2-\tau)=\mu_1
\elabel{mu_one_expo}
\end{equation}
If $\mu_1$ is known exactly ($\mu_1=2$ for bulk driving Manna, Oslo and
Abelian BTW Models, $\mu_1=1$ for boundary drive), then \Eref{mu_one_expo} gives
rise to a
\keyword{scaling relation}. Normally, there are no further, strict scaling
relations. However, the assumption of narrow joint
distributions suggests $D(\tau-1)=z(\alpha-1)$ \etc
\citep{ChristensenFogedbyJensen:1991,ChessaETAL:1999,JensenPruessner:2004}.
If the exponent $\mu_1$ is given by a mathematical identity and
$\ave{s}$ serves as an analytically known reference in the numerical
simulation, then $\mu_1$ should not
feature in the numerical analysis to extract the scaling exponents $D$ and $\tau$.
Rather, when fitting $\mu_n$ versus $D(1+n-\tau)$, this should be
replaced by $D(n-1)+\mu_1$.

Fitting $\mu_n$ in a linear fit (without corrections) against
$D(1+n-\tau)$ (or against $D(n-1)+\mu_1$ if $\mu_1$ is known exactly) 
is, in fact, the third step in the procedure described in this section. In principle, the
$n>\tau-1$ do not need to be integer valued. They have to be large
enough to avoid a significant corrections due to the lower cutoff, and
small enough to keep the relative statistical error small. Non-integer
$n$ can be computationally expensive, as they normally require at least
one library call
of \code{pow}. 

While each
estimate $\mu_n$ for every $n$ should be based on the entire ensemble,
considering them together in the same fit to extract the exponents $D$
and $\tau$ introduces correlations, which are very often unaccounted for. As a
result both goodness-of-fit as well as the statistical error for the
exponents extracted are (unrealistically) small. 

There are a number of strategies to address this problem. The simplest
is to  up-scale the error of the $\mu_n$ as if every estimate was based
on a separate, independent set of raw data. Considering $M$ moments
simultaneously, their error therefore has to scaled up by a factor
$\sqrt{M}$ \citep{HuynhPruessnerChew:2011}. In a more sophisticated approach, one may extract estimates
from a series of sub-samples \citep{Efron:1982,Berg:1992,Berg:2004}.

It often pays to go through the process of extracting the exponents $D$
and $\tau$ at an early stage of a simulation campaign, to identify
potential problems in the data. Typical problems to watch out for
include
\begin{itemize}
\item Corrections are too strong because system sizes are too small.
\item Results are too noisy because sample sizes are too small, often
because the
system sizes are too big.
\item Results have so little noise that fitting functions need to
contain too many free parameters.
\item Too few data points (\ie too few different system sizes $L$ or
different moments $n$).
\item Large event sizes suffer from integer overflow, resulting in
seemingly negative or very
small event sizes.
\item Data identical in supposedly different runs, because of using the
same seed for the random number generator.
\item Transients chosen too short.
\end{itemize}

\subsection{Statistical errors from chunks}
\slabel{chunks}
One of the key-ingredients in the procedures described above is a
reliable estimate for the statistical error of the estimates of the
individual moments. The traditional approach is to estimate the
variance, $\sigma^2(s^n)=\ave{s^{2n}}-\ave{s^n}^2$ of each moment, so
that the statistical error of the estimate of $\ave{s^n}$ is estimated
by $\sigma^2(s^n)/\sqrt{N/(2\tau+1)}$, where $N/(2\tau+1)$ is the number
of effectively independent elements in the sample with correlation time
$\tau$.

This approach has a number of significant drawbacks. Firstly, each
moment $\ave{s^n}$ requires a second moment, $\ave{s^{2n}}$, to be
estimated as well. Considering a range of moments, this might
(almost) double the computational effort. Rather dissatisfyingly, the highest
moment estimated itself cannot be used to extract its finite size
scaling exponent
$\mu_n$, because its variance is not estimated. Furthermore, because of their very high
powers, the moments entering the estimates of the variances and thus the
variances themselves have large statistical errors and are prone to
integer overflow. 

Estimating the effective number of independent elements in the sample is
a hurdle that can be very difficult to overcome. Usually, it is based on
an estimate of the correlation time $\tau$. If $\ave{s_i
s_j}-\ave{s}^2=\sigma^2(s) \exp(-|i-j|/\tau)$, then the variance of the estimator
\begin{equation}
\overline{s}=\frac{1}{N}\sum_i^N s_i
\end{equation}
of $\ave{s}$ for $N\gg\tau$ is 
\begin{equation}
\sigma^2(\overline{s}) = \frac{1}{N^2} \sum_{ij}^N \left(\ave{s_i s_j} -
\ave{s}^2\right) \approx \frac{1+\exp(-1/\tau)}{N(1-\exp(-1/\tau))}
\sigma^2(s) \approx \frac{2\tau+1}{N} \sigma^2(s) 
\end{equation}
as if the sample contained only $N/(2\tau+1)$ independent elements. 

The main difficulty of this strategy is a reliable estimate of $\tau$
which often cannot be easily extracted from $\ave{s_i
s_j}-\ave{s}^2$ because of noise and the presence of other exponential
contributions, of which $\exp(-|i-j|/\tau)$ is the slowest decaying one.
Moreover, in principle $\tau$ has to be measured for each
observable separately (even when it makes physically most sense to
assume that the system is characterised by a single correlation time).

To avoid these difficulties, one may resort to a simple sub-sampling
plan. As discussed below (also \Sref{output}), it is a matter of mere convenience and
efficiency to repeatedly write estimates of moments based on a
comparatively small sample into the output stream of a simulation and reset the
cumulating variables. In the following these raw estimates based on a
small sample are referred to as \keyword{chunks}.
If their sample size is significantly larger than
the correlation time, then each of these estimates can be considered as
independent and the overall estimates based on it has its statistical
error estimated accordingly. For example, if $m_i$ with $i=1,2,\ldots,M$
are estimates of $\ave{s^n}$ all based on samples of the same size $N$, say
$m_i=\sum_{j}^N s_{ij}^n$ with $s_{ij}$ the $j$th element of the $i$
sample, then the overall unbiased and consistent estimator
\citep{Brandt:1998} of $\ave{s}$ is 
\begin{equation}
\overline{m}=\frac{1}{M} \sum_i^M m_i
\end{equation}
which has an estimated standard deviation of
$(\overline{m^2}-\overline{m}^2)/(M-1)$ where
\begin{equation}
\overline{m^2}=\frac{1}{M} \sum_i^M m_i^2 \ .
\end{equation}

One crucial assumption above is that the $m_i$ are independent, which
can always be achieved by merging samples. As long as $M$ remains
sufficiently large, one may be generous with the (effective) size of
the individual samples \citep{FlyvbjergPetersen:1989}.

\begin{figure}
\begin{center}
\includegraphics[width=0.7\linewidth]{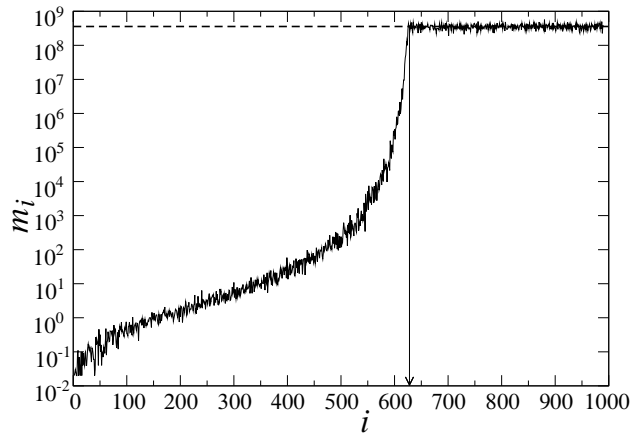}
\end{center}
\caption{\flabel{transient}
Example of the transient behaviour of an observable (here the
average avalanche size in the one-dimensional Manna Model with $L=65536$) as a
function of the chunk index in a log-lin plot \citep[data
from][]{HuynhPruessnerChew:2011}. The straight dashed line shows the exact
expected average $\ave{s}$, \Eref{exact_aves}. The  arrow indicates the
chunk from where on stationarity is roughly reached. A generous multiple
of that time should be taken as the number of chunks to discard in order
to ensure that correlations (and thus dependence on the initial setup)
are essentially overcome.
}
\end{figure}

Chunks also allow a more flexible approach to determining and discarding
transient behaviour from the sample supposedly taken in the stationary
state. The transient can be determined as a (generous) multiple of the
time after which (ideally all or several) observables no longer deviate more from the asymptotic
or long time average than their characteristic variance. Where
observables are known exactly \citep[e.g. the average avalanche
size][]{Pruessner_aves:2012}, they can be used as a suitable reference.
\Fref{transient} shows the transient behaviour of the average avalanche
size in a realisation of the Manna Model. A more cautious strategy is
to consider a series of different transients and study the change in the
final estimates (with their estimated error) as a function of the
transient discarded.

\section{Algorithms and data organisation}
In the following, a range of numerical and computational procedures are
discussed that are commonly used in the numerical implementation of SOC
models \citep[for a more extensive review see][]{Pruessner:2012:Book}. Some of them are a matter
of common sense and should be part of the coding repertoire of every
computational physicist. However, it is not always entirely obvious how
these ``standard tricks'' are used for SOC models.

In the following, the focus is on computational performance, which often
comes with the price of lower maintainability of the code. The amount of
real time spent on writing code and gained by making it efficient,
should account for the time spent on debugging and maintaining it. 

Most
of the discussion below is limited to algorithmic improvements. The aim
is produce code that communicates only minimally with the ``outside
world'', because in 
general, interaction with the operating system, as required for writing
to a file,
is computationally expensive and
extremely slow. The UN*X operating system family (including, say, Linux
and Mac OS X) distinguishes two different ``modes'' by which an executable keeps
the CPU busy: By spending time on the (operating) system and by spending it in
``user mode''. Roughly speaking, the former accounts for any interaction
between processes, with external controls or peripherals, including 
writing files. The latter accounts for the computation that takes place
solely on the CPU (ALU, FPU, GPU, \etc) and the attached RAM. Tools like \code{time} and
library functions like \code{getrusage} provide an interface to assess
the amount of various resources used, while being themselves or
resulting in systems calls.

Of course, the literature of computational physics in general is vast.
Reviews and texts that are of particular use in the present context
include
\citet{KernighanRitchie:1988,CormenLeisersonRivest:1996,Knuth:1997Vall,NewmanBarkema:1999,Berg:2004,LandauBinder:2005,PressETAL:2007}.

\subsection{Stacks}
\slabel{stack_discussion}
The definition of most SOC models makes no reference to the method to
identify active sites, \ie sites that are due to be updated. In
principle, an implementation of an SOC model could therefore repeatedly
scan the entire lattice to find the relevant sites. This is, however,
very inefficient and therefore should be avoided. Instead, the index of
active sites (or their coordinates) should be organised in a list. Every
site in that list is subsequently updated. Moreover, it is often very
important to know whether a site is maintained in the list or not.
Sometimes this can be determined implicitly (for example, when a site is
guaranteed to reside on the list from the moment its height exceeds the
threshold), sometimes this is done explicitly by means of a flag
associated with the site. The following contains a more detailed
discussion of the various techniques available.

The most commonly used form of a list is a \keyword{stack}, called so,
because this is how it appears to be organised. It consists of a vector,
say \code{int stack[STACK\_SIZE]}, of pre-defined size
\code{STACK\_SIZE}. It must be large enough to accommodate the maximum
number of simultaneously active sites. Simulating large lattices, a
balance has to be struck between what is theoretically possible and what
is happening in practise.

The type of the stack, vector of \code{int} in the example above, is
determined by the type it is meant to hold. If it holds the index of
active sites, it is likely to be be \code{int}, but it may also hold
more complex objects, say, coordinates of active particles (but see
below). The number
of objects currently held by the stack is stored in \code{int stack\_height}.

If \code{STACK\_SIZE} is
smaller than the theoretical maximum of active sites, \code{int stack\_height} 
has to be monitored as to prevent it from exceeding
\code{STACK\_SIZE}. The outcome of the simulation is undefined if that
happens, because the exact position in memory of
\code{stack[STACK\_SIZE]} is \latin{a priori} unknown. If therefore
\code{stack\_height} exceeds \code{STACK\_SIZE}, memory has to be 
extended one way or another. For example, one may use \code{realloc()},
which assumes, however, that enough memory is actually available. Modern
operating systems all provide virtual memory which is transparently
supplemented by a swap file residing on the (comparatively slow) hard
drive. This is to be avoided because of the computational costs
associated. It may thus pay off for the process itself to make
use, temporarily, of a file to store active sites. The alternative to
abandon the particular realisation of the simulation introduces a bias
away from rare events which is likely to have significant effect on
observables. The same applies obviously if activity is suppressed if it
reaches the maximum level.

There are two fundamental operations defined on a stack,
\begin{lstlisting}
#define PUSH(a) stack[stack_height++]=(a)
#define POP(a)  (a)=stack[--stack_height]
\end{lstlisting}
where \code{PUSH(a)} places \code{(a)} on the stack and \code{POP} takes
an element off. The underlying idea is literally that of a stack: When a site
becomes active, its index goes on a pile (\code{PUSH}) so that each index number on that
pile represents a site waiting to be updated. When that happens, it is
removed from the pile (\code{POP}).

It simplifies the code greatly if all objects on the stack are, in a
sense, equivalent. For example, all sites on a stack are active.
Guaranteeing this is not necessarily trivial, because the manipulation
of one item on the stack may affect the state (and thus the
eligibility) of another item on the stack. It is therefore advisable
to ensure that all elements on the stack are distinct. In SOC models
that means that active sites enter the stack exactly once, namely when
the \emph{turn} active. If an active site is charged again by a toppling
neighbour, a new copy of its index is \emph{not} placed on the stack. In
the Manna Model, for instance, the single line of code to place objects
on the stack could be
\begin{lstlisting}
if (z[i]++==1) {PUSH(i);}
\end{lstlisting}
so that the index \code{i} of a site enters when it is charged while
its height $z$ is \emph{at} the critical value $z^c$. The line should
\emph{not} read \code{if (z[i]++>=1) {PUSH(i);}}.

Unfortunately, the very data structure of a stack, which in the present
context may better be called a \keyword{LIFO} (last in, first out),
suggests a particular procedure to explore active sites, namely a
\abbrevdef{depth first search}{DFS}; Whenever a toppling site activates
its neighbours, one of them will be taken off first by the next call of
\code{POP}, toppling in turn. Activity thus spreads very far very
quickly, then returning, then spreading far again, rather than
``burning locally''. In fact, in DS-FFM a DFS is probably the simplest
way of exploring a cluster of trees.

The alternative, a \abbrevdef{breadth first search}{BFS} requires
slightly greater
computational effort because it normally makes use of a \keyword{FIFO}
(first in, first out). The last object to arrive on a FIFO is the last
one to be taken off, exactly the opposite order compared to a stack.
Naively, this may be implemented by removing items from the front,
\code{stack[0]}, and using \code{memmove()}\footnote{Dedicated library functions
like \code{memmove} and \code{memcpy} are generally much faster than naive procedures
based on loops, although the latter can be subject to significant
optimisation by the compiler.} to feed it from the end, lowering \code{stack\_height}.
This approach, however, is computationally comparatively costly.
A faster approach is to organise the stack in a queue,
organised in a ring (circular buffer) to keep it finite, where a string
of valid data grows
at the end while retreating from the front.

In Abelian models, where the statistics of static features of
avalanches, such as size and area, do not depend on the details of the
microscopic dynamics\footnote{But note the strict definition of
Abelianness discussed on \pref{microscopic_Abelianness}.}, working
through the stack using \code{POP} may be acceptable. Where temporal
features are of interest too, the microscopic dynamics must implement a
suitable microscopic time scale. Often the microscopic
timescale is given by Poissonian updates, for example by
active sites toppling with a Poissonian unit rate. 

In principle that means that waiting times between events (sites
toppling) are themselves random variables. If a faithful representation
of the microscopic time is desired, then the random waiting times can be
generated by taking the negative logarithm of a random number drawn from
a uniform distribution on $(0,1]$. If an approximate representation of
the Poisson processes is acceptable \citep[which, in fact converges to the
exact behaviour in the limit of large numbers of active
sites, see][]{Liggett:2005}, then elements are taken off the stack at
random and time is made to progress in steps of
\code{1./stack\_height}. If \code{stack\_height} remains roughly constant,
than on average \code{stack\_height} events occur per unit time as
expected in a Poisson process. A simple implementation reads
\begin{lstlisting}
int rs_pos;
#define RANDOM_POP(a) rs_pos=rand() % stack_height; (a)=stack[rs_pos]; POP(stack[rs_pos])
\end{lstlisting}
where the last operation, \code{POP(stack[rs\_pos])} overwrites the
content of \code{stack[rs\_pos]} by \code{stack[stack\_height-1]}
decrementing \code{stack\_height} at the same time. When selecting the
random position on the stack via \code{rs\_pos=rand() \% stack\_height} a
random number generator has to be used (\Sref{RNG}), which only for
illustrative purposes is called \code{rand()} here.

One consequence of the constraint of distinct objects on the stack is
that a site may need to topple several times before being allowed to
leave the stack. In Abelian models some authors circumvent that by
placing a copy of the site index on the stack every time a pair of particles has to be
toppled from it, which can be implemented easily by removing an appropriate
number of particles from the site each time it enters the stack. As a
result, however, stacks may become much larger, \ie a greater amount of
memory has to be allocated to accommodate them.

Depending on the details of the microscopic dynamics, an possible alternative is to relax a
site completely after it has been taken off the stack, for example in the Manna
Model:
\begin{lstlisting}
while (stack_height) {
  RANDOM_POP(i);
  do {
    topple(i);        /* Site i topples, removing two particles from i. */
    avalanche_size++; /* avalanche_size counts the number of topplings. */
  } while (z[i]>1);
}
\end{lstlisting}
where \code{topple(i)} reduces \code{z[i]} by $2$ each time. If the 
avalanche size counts the number of topplings performed,
\code{avalanche\_size} has to be incremented within the loop. Counting
only \emph{complete} relaxations would spoil the correspondence with
exact results.

An alternative approach with different microscopic time scale is to
topple a site on the stack only once, and take it off only once it is
fully relaxed. This approach requires some ``tempering'' with the stack:
\begin{lstlisting}
while (stack_height) {
  i=rand() % stack_height;
  topple(stack[i]);
  if (z[i]<=1) POP(stack[i]);
}
\end{lstlisting}

In systems with parallel update, where all sites at the beginning of a
time step have to be updated concurrently before updating the generation
of sites that have been newly activated, a red-black approach
\citep{DowdSeverance:1998} can
be adopted. This requires the use of two stacks, which have to be
swapped after completing one:
\begin{lstlisting}
int *stack, stack_height=0;
int rb_stack[2][STACK_SIZE], next_stack_height;
int current_stack, next_stack;

#define NEXT_PUSH(a) rb_stack[next_stack][next_stack_height++]=(a)
#define NEXT_POP(a)  (a)=rb_stack[next_stack][--next_stack_height]

...
current_stack=0;
next_stack=1;
stack=rb_stack[current_stack];
...
PUSH(i);
...
for (;;) {
  while (stack_height) {
    ...
    POP(i);
    ...
    NEXT_PUSH(j);
    ...
  }
  if (next_stack_height==0) break;
  /* Swap stacks. */
  stack_height=next_stack_height;
  next_stack_height=0;
  current_stack=next_stack;
  stack=rb_stack[current_stack];
  next_stack=1-next_stack;
}
/* Both stacks are empty. */

\end{lstlisting}
The use of the pointer \code{stack} is solely for being able to use the macros
\code{PUSH} and \code{POP} defined earlier. Otherwise, it might be more
suitable to define macros \code{CURRENT\_PUSH} and \code{CURRENT\_POP}
corresponding to \code{NEXT\_PUSH} and \code{NEXT\_POP}.

A stack should also be used when determining the area of an avalanche,
\ie the number of distinct sites toppled (or visited, \ie charged). To mark each site that has
toppled during an avalanche and to avoid double counting, a flag has to
be set, say \code{visited[i]=1} or \code{site[i].visited=1} (see
\Sref{sites_and_neighbours}). Counting how often the flag has been newly
visited then gives the avalanche area. However, in preparation for the
next avalanche, the flags have to be reset. This is when a stack comes
handy, say
\begin{lstlisting}
int area_stack[SYSTEM_SIZE];
int area_stack_height=0;
#define AREA_PUSH(a) area_stack[area_stack_height++]=(a)
#define AREA_POP(a)  (a)=area_stack[--area_stack_height]
...
/* For each toppling site. */
if (visited[i]==0) {
  visited[i]=1;
  AREA_PUSH(i);
}
...
/* After the avalanche has terminated. 
 * area_stack_height is the avalanche area. */
...
/* Re-initialise */
while (area_stack_height) {
  AREA_POP(i);
  visited[i]=0;
}
...
\end{lstlisting}
In the example above, the area is tracked implicitly in
\code{area_stack_height}. The re-initialisation can be further improved
using \code{while (area_stack_height) visited[area_stack[--area_stack_height]]=0}.

\subsection{Sites and Neighbours}
\slabel{sites_and_neighbours}
In SOC models, every site has a number of properties, most importantly
the local degree of freedom, but also (statistical) observables which are
being measured and updated as the simulation progresses. Other
information associated with each site are flags (such as the one
mentioned above to indicate whether a site had been visited) and even
the neighbourhood (discussed below). In fact, the site itself may be
seen as the \keyword{key} associated with all that information. That key
might represent information in its own right, say, the coordinate, it
might be an index of a vector, or a pointer.

\subsubsection{Pointers and structures}
A word of caution is in order with regard to pointers. The programming
language C lends itself naturally to the use of pointers. However, code
on the basis of pointers is difficult to optimise automatically at
compile time. Depending on the quality of the compiler and the
coding an index based
implementation (which is also more portable) may thus results in faster
code than the seemingly more sophisticated implementation based on
pointers. 

That said, in theory placing pointers on the stack, which gives
immediately access to a relevant object should be faster than using
indices, which are effectively an offset relative to a base:
\code{b=z[stack[i]]} might result in machine code of the form
\code{b=*(z+*(stack+i))} which contains one more addition than
\code{b=*stack[i]} resulting in \code{b=**(stack+i)} if \code{stack} is a
vector of pointers.

Similar considerations enter when using structures, which provide very
convenient and efficient ways of organising and encapsulating data 
associated with each site. For example
\begin{lstlisting}
struct site_struct {
  int height;
  char visited;
};
\end{lstlisting}
defines a structure with two members, \code{height} and \code{visited}.
Declaring a variable \code{struct site\_struct site[10]} allows the
individual elements to be accessed in a structured way, say
\code{site[i].height++, site[i].visited=1}. There are a number of
computational drawbacks, which are, however, normally outweighed by the
better maintainability of the code. 
\begin{itemize}
\item Depending on the platform and the compiler, padding might become
necessary, \ie some empty space is added to the structure (\Sref{neighbourhood_information}, 
\pref{padding}). The memory
requirements of the structure is thus greater than the memory
requirements for each variable when defined individually.
\item Again depending on the platform as well as the compiler, without 
padding some operations on some types may require more CPU cycles (in
particular when floating point types are used).
\item Members within the structure are accessed similar to elements in a
vector, namely by adding an offset. Access to the first member (where
no offset is needed, \code{site[i].height} in the example above) can thus 
be faster than access to the other
members (\code{site[i].visited} above). Because of that additional
addition, the approach is often slower than using separate vectors for
each member of the structure.
\end{itemize}

\subsubsection{Neighbourhood information}
\slabel{neighbourhood_information}
It can be convenient, in particular for complicated topologies or when
the neighbourhood information is computed or supplied externally, to store
information 
about the local neighbourhood in a site structure, for example:
\begin{lstlisting}
struct site_struct {
  ...
  int neighbour[MAX_NEIGHBOURS];
  int num_neighbours;
};
\end{lstlisting}
Because of the
significant memory requirements, this is often not viable for large
lattices. Again, instead of addressing neighbours by their index, pointers can be
used, which often produces very efficient and elegant code.

The neighbours of each site thus are calculated and stored at
the site only once. The strategy of pre-calculated neighbourhoods goes
back to the very beginning of computational physics, when access to
memory was much faster than doing such calculations
on-the-fly.\footnote{Back in the days when lookup tables for modulo operations were in fashion.} This,
however, has changed. It can be \emph{much} faster to determine a
neighbourhood on-the-fly than looking it up, unless, of course, the
topology is so complicated that it becomes computationally too costly.
Unfortunately, it is often difficult to try out different
implementations (lookup tables and calculation on the fly), as the setup
of a neighbourhood is at the heart of a lattice simulation.

As for calculating neighbourhoods, in one dimension the index of a site,
which is strictly only a key to access the information, is often
associated with its position on a one-dimensional lattice. Actual
computation takes place only at boundaries. If the right neighbour of
site \code{i} in the bulk is \code{i+1}, it may not exist on the right boundary or
be \code{0} if \abbrevdef{periodic boundary conditions}{PBC} apply in an
implementation in C where the index of a vector of size \code{LENGTH}
can take values from \code{0} to \code{LENGTH-1}. Similarly, the left
neighbour is \code{i-1} in the bulk and \code{LENGTH-1} at \code{i=0} in case of
periodic boundaries. Those are most easily implemented in the form
\code{left=(i+LENGTH-1)\%LENGTH} and
\code{right=(i+1)\%LENGTH} respectively using a modulo operation. The
shift by \code{LENGTH} in the former avoids problems with negative
indices at \code{i=0}.

A less elegant but often faster implementation is to determine whether a
site is at the boundary before assigning the value for the neighbour,
such as
\begin{lstlisting}
if (i==0) left=LENGTH-1;
else left=i-1;
\end{lstlisting}
or just \code{left=(i==0)?LENGTH-1:i-1}, which is more readable.
This method is also more flexible with respect to the boundary condition
implemented. Reflecting boundary conditions, for example are implemented
by \code{left=(i==0) ? 1 : i-1}. Open boundary conditions, on the other
hand, might require special attention. If possible, they are best
implemented using \plabel{padding}\keyword{padding}, \ie by pretending that a
neighbouring site exists, which, however, cannot interact with the rest
of the lattice, for example, by making sure that it never fulfils the
criterion to enter the stack. Such a site may need to be ``flushed''
occasionally to prevent it, for example, from fulfilling the criterion due to integer
overflow. One might either assign one special site, say the variable \code{dump} in
\code{left=(i==0)? dump : i-1} or allocate memory for \code{LENGTH+2} sites with an index from \code{0} to
\code{LENGTH+1}, with valid sites ranging from \code{1} to \code{LENGTH}
with sites \code{0} and \code{LENGTH+1} receiving charges without
toppling in turn. This procedure also allows a very efficient way to
determine the number of particles leaving the system, the \keyword{drop
number} \citep{KadanoffETAL:1989}.

Usually only
in higher dimensions, one distinguishes reflecting boundary
conditions, where the particle offloaded is moved to another site
(normally the mirror image of the ``missing'' site), and
``closed'' boundary conditions, where the number of nearest neighbours is
reduced and shed particles are evenly re-distributed among them.

Most of the above techniques remain valid in higher dimension, where the data
can be organised in either a one-dimensional vector or a
multidimensional vector. The former strategy makes use of macros of the
form
\begin{lstlisting}
#define COORDINATE2INDEX(x,y,z)   ((x)+(LENGTH_X*((y)+LENGTH_Y*(z))))
#define INDEX2COORDINATE(i,x,y,z) z=(i)/(LENGTH_X*LENGTH_Y),y=((i)/LENGTH_X)%LENGTH_Y,x=(i)%LENGTH_X
\end{lstlisting}
The use of the coma operator in the second macro helps to avoid errors
when omitting curly brackets in expressions like
\code{if (1) INDEX2COORDINATE(i,x,y,z);}. Where stacks are used to hold
coordinates, the multiple assignments needed to store and fetch all of them may
computationally outweigh the benefit of not having to calculate
coordinates based on a single index.

The two biggest problem with the use of multi-dimensional vectors is
their ambiguity when used with fewer indices and the consistency when
passing them to functions. Both subtleties arise because of the logical
difference between a vector of pointers to a type and the interpretation
of a lower-dimensional
variant of a multi-dimensional vector. While C makes that distinction,
there is no syntactical difference between the two. For example
\begin{lstlisting}
int a[2][10];

a[0][5]=7;
\end{lstlisting}
is a multi-dimensional vector using up \code{2*10*sizeof(int)}
sequential bytes of
memory. Each \code{a[i]} is the starting address of
each row $i=0,1$. On the other hand
\begin{lstlisting}
int *a[2];
int row1[10], row2[10];
a[0]=row1; a[1]=row2;

a[0][5]=7;
\end{lstlisting}
makes \code{a} a vector of pointers, using up \code{2*sizeof(* int)}
bytes of memory, while each row uses \code{10*sizeof(int)} bytes.
Both snippets of code
declare \code{a} to be completely different objects, yet, for all
intents and purposes in both cases \code{a} will behave like a two-dimensional array. 
That
is, until it is to be passed as an argument to another function. In the
first case, that function can be declared by \code{function(int array[2][10])}, informing it about the dimensions of the array, and
subsequently called using \code{function(a)}. The two-dimensional vector
\code{a} will behave as in the calling function. In fact, the
function will even accept any other vector, lower dimensional or not,
passed on to it as an argument (even when the compiler may complain).

In the second case, \code{a} is a vector of pointers to \code{int}, and
so a function taking it as an argument must be declared in the form
\code{function(int **a)}, using additional arguments or global constants
(or variables) to inform it about
the size of the vector. The two versions of the functions are
incompatible, because a two-dimensional vector is really a
one-dimensional vector with a particularly convenient way of addressing
its components. 
In particular, the two-dimensional vector cannot be
passed to the function designed for the second case using, say,
\code{function(\&a)} or \code{function((int **)a)}.

While these issues normally are resolved at the time of coding they
can cause considerable problems when the memory allocation mechanism for
the vector is changed. This happens, in particular, when lattice sizes
are increased
during the course of a simulation campaign. Initially, one might be tempted to
define a lattice globally (stored in BSS or data segment) or as automatic variables taken from the stack,
choosing a multi-dimensional array for convenience. 
Later on, they make be taken from the (usually much bigger) heap using \code{malloc()}, at
which point the way they are accessed may have to be changed.
The latter approach is the most flexible but possibly not the most
convenient way of allocating memory for large items.

Finally, it is advisable to scan sites (when sweeping the lattice is
unavoidable or
scanning through a local neighbourhood) in a way that is local in memory
and thus cache. The first option, declaring a two-dimensional vector in
a single step, makes that more feasible than the second option, where different rows might end
up at very different regions of memory. Not using higher dimensional
vectors at all, however, is probably the best performing option.

\subsection{Floating Point Precision}
\slabel{numerical_precision} 
Very little and at times too little attention is being paid to the
effect of limited floating point precision. Most SOC models can be
implemented fully in integers even when their degrees of freedom are
meant to be real valued, such as the Zhang Model \citep{Zhang:1989}, the
Bak-Sneppen Model \citep{BakSneppen:1993} or the Olami-Feder-Christensen
Model Model \citep{OlamiFederChristensen:1992}.  In case of the latter,
floating point precision has been found to significantly affect the
results \citep{Drossel:2002}.

Where \keyword{random floating point} numbers are drawn, they might in fact contain
much fewer random bits than
suggested by the size of their mantissa. In that case, an implementation in integers is
often not only faster but also ``more honest''. Where rescaling of
variables cannot be avoided and occurs frequently, multiplying by a
constant inverse often produces faster code than division.

Over the last decade or so, the floating point capabilities of most
common CPUs have improved so much, however, that the difference in
computational costs between integers and floating point arithmetics is
either negligible or not clear-cut. The most significant disadvantage of the latter is the
limited control of precision that is available on many platforms. 

The levels of precision as defined in the IEEE standard 754 that are
very widely used are single, double and extended. They refer to the
number of bits in the mantissa determined when floating point operations
are executed, \ie they are the precision of the \abbrevdef{floating
point unit}{FPU}. The precision the FPU is running at depends on
platform, environment, compiler, compiler switches and the program
itself. Some operating systems offer an IEEE interface, such as
\code{fpsetprec()} on FreeBSD, and \code{fenv} on Linux.

Results of floating point arithmetics are stored in variables that may not offer the
same level of precision the FPU is running at and in fact it is possible that none of the data types
available matches a particular level of precision set on the FPU.
Crucially, the precision setting of the FPU normally affects \emph{all}
floating point operations on \emph{all} floating point variables, regardless of
type, \eg information is lost when results are calculated with extended
precision and stored in variables offering only single precision.
A notorious error observed on systems which default to extended
precision, in particular Linux on x86, occurs when comparisons between
variables produce different outcomes depending on the position in the
code --- at one point the result calculated may still reside on the FPU and thus
offer extended precision, whereas at a later point the result is
truncated after being written to memory. This can lead to serious
inconsistencies when data is held in an ordered tree. Compiler switches
like \code{-ffloat-store} for \code{gcc} help in these cases.

The commonly used \code{gcc} compiler offers three basic floating point
types, \code{float}, \code{double} and \code{long double}, matching the
three levels of precision mentioned above. The very nature of SOC means
that observables span very many order of magnitudes. If variables that
accumulate results, such as moments, are too small (\ie have a mantissa
that is too small), smaller events may not accumulate at all any
more once the variable has reached a sufficiently large value. This can
skew estimates considerably where very large events occur very rarely.
The
macros \code{FLT\_EPSILON}, \code{DBL\_EPSILON} and \code{LDBL\_EPSILON} in
\code{float.h} give a suggestion of the relative scale of the problem.
It can be mitigated by frequently ``flushing'' accumulating variables
(see \Sref{output}).

\subsection{Random Number Generators}
\slabel{RNG}
\abbrevdef{Random Number Generators}{RNGs} are a key ingredient in many
areas of computational physics, in particular in Monte-Carlo
and Molecular Dynamics simulations. The vast majority of them, strictly,
are not random, but follow instead a deterministic but convoluted computational
path. RNGs are constantly being improved and evaluated, not least because of
their use in cryptography. An introduction into the features of a good
RNG can be found in the well-known Numerical Recipes
\citep{PressETAL:2007},
with further details to be found in the review by \citet{Gentle:1998}.

A ``good'' random number generator is one that offers a
reasonable compromise between two opposing demands, namely that of speed
and that of quality. In most stochastic SOC models, the RNG 
is used \emph{very} often and thus typically consumes about
half of the overall CPU time. Improving the RNG is thus a particularly
simple way of improving the performance of an implementation. Because
the variance (square of the standard deviation) of an estimate vanishes inversely
proportional with the sample size it is based on, the performance of an implementation
is best measured as the product of variance and CPU time spent ``for
it''.
However, one is ill-advised to cut corners by using a very fast RNG
which has statistical flaws. The resulting problem may be very subtle
and might not show until after a very detailed analysis. 

One of the problems is the period of an RNG. Because RNGs generally have
a finite state, they are bound to repeat a sequence of random numbers
after a sufficient number of calls,
at which point the simulation using the random numbers produces only
copies of previous results. With
improving hardware the RNG must therefore be re-assessed. A ``good RNG''
is a function of time, and very much a function of perception, as a
mediocre RNG might appear to be a fantastic improvement over a poor RNG.
It is good
practise to use more than one random number generator to derive the same
estimates and compare the results. 

The C library's implementation of \code{rand()} is legendary for being
unreliable and can be very poor. At the very least, it is essentially
uncontrolled, although, of course, standards exist, which are, however,
not always adhered to. It is fair to say that pure linear congruential RNGs
are somewhat (out-)dated and indeed rarely used. They are, however,
sometimes combined or enhanced with more sophisticated techniques. In
recent years, the Mersenne Twister \citep{MatsumotoNishimura:1998a,Matsumoto:2008} has become very widely used,
yet, criticised by \citet{Marsaglia:2005} who proposed in turn
KISS \citep[][but see \citealp{Rose:2011}]{Marsaglia:1999}, which is a remarkably simple RNG. The GNU Scientific
Library \citep{GalassiETAL:2009} contains an excellent collection of random number
generators.

Somewhat more specific to the use of RNGs in SOC models is the frequent
demand for random bits, for example in order to decide about the
direction a particle is taking. Because every acceptable RNG is made up
of equally random bits, each and everyone of them should be used for
random booleans. These bits can be extracted one-by-one, 
by bit-shifting the random integer or
by shifting a
mask across, as in 
\begin{lstlisting}
#define RNG_MT_BITS (32)
#define RNG_TYPE unsigned long
RNG_TYPE mt_bool_rand=0UL;
RNG_TYPE mt_bool_mask=1UL<<(RNG_MT_BITS-1);
#define RNG_MT_BOOLEAN ( ( mt_bool_mask==(1UL<<(RNG_MT_BITS-1)) ) ? ((mt_bool_mask=1UL, mt_bool_rand=genrand_int32()) & mt_bool_mask) : (mt_bool_rand & (mt_bool_mask+=mt_bool_mask)) )
\end{lstlisting}
based on the Mersenne Twister.
In general, bit shifts to the
left using \code{a+=a} instead of \code{a<<=1} are faster, because the
latter requires one more CPU cycle to write the constant \code{1} into
the CPU's register.

More generally, integer random numbers have to be chosen uniformly from
the range $\{0,1,\ldots,n-1\}$ suggesting the use of the modulo
operation, \code{r=rand()\%n}. However, if \code{rand()} produces random
integers uniformly from $0$ up to and including \code{RAND\_MAX}, then the
modulo operation skews the frequencies with which random number occurs 
towards smaller values if \code{RAND\_MAX+1} is not an integer multiple 
of $n$. The effect is of order $n/$\code{(RAND\_MAX+1)} and thus is
negligible if $n$ is significantly smaller than \code{RAND\_MAX}.
However, picking a site at random on a very large lattice or an element
from a very large stack, this effects
becomes a realistic concern. In that case, the modulo operation can be
used on a random number drawn uniformly among integers from $0$ up to
and including $R-1$, where $R$ is a multiple of $n$ and ideally the
largest multiple of $n$ less or equal to \code{RAND\_MAX+1}:
\begin{lstlisting}
const long long int n=...;
/* The constant multiple_minus_1 is made to have type as the return
 * value of rand(). */
const int multiple_minus_1=(n*((((long long int)RAND_MAX) + 1LL)/n))-1LL;
int r;
#define RANDOM(a) while ((r=rand())>multiple_minus_1); (a)=r%n 
\end{lstlisting}
where \code{multiple\_minus\_1} plays the r{\^o}le of $R-1$.
When determining the maximum multiple, it is crucial that the operation
\code{RAND\_MAX+1} is performed using a type where the addition does not lead to
rounding or integer overflow. The latter is also the reason why
one is subtracted in the expression for \code{multiple\_minus\_1}, which
otherwise might not be representable in the same type as the return
value of \code{rand()}, which is necessary to avoid any unwanted type
casting at run time.\footnote{This is one of the many good reasons to
use constants rather than macros  \citep{VanderLinden:1994,KernighanPike:2002}.}

The initial seed of the RNG needs to be part of the
output of the programme it is used in, so that the precise sequence of events can be reproduced in case
an error occurs. Some authors suggest that the initial seed itself
should be random, based, for example, on \code{/dev/random}, or the
library functions \code{time()} or \code{clock()},\footnote{Both
functions are bad choices on clusters where several instances of the
same programme are intended to
run in parallel. The function \code{time()} changes too slowly
(returning the UN*X epoch time in seconds) and the function
\code{clock()} wraps after about $36$ minutes, so that neither function
guarantees unique seeds. In general, seeding is best done explicitly.}
and that the RNG
carries out a ``warm-up-cycle'' of a few million calls
\citep{Jones:2010}. After that, it
is sometimes argued, chances are that one sequence of (pseudo) random numbers is
independent from another sequence of random numbers generated by the
same RNG based on a different seed. Fortunately,
some RNGs, in particular those designed for use on parallel machines,
offer a facility to generate sequences that are guaranteed to be
independent. 
Where poor-man's parallel computing (many instances of the
same simulation running with different seeds) takes place, independent
sequences are of much greater concern than in situations where different
parameter settings are used in different instances. In the former case
the data of all instances will be processed as a whole, probably under
the assumption that it is actually independent. 
In the latter case, the results will enter differently and using even an
identical sequence of random numbers will probably not have a noticeable
effect.
All these caveats are
put in perspective by the fact that most SOC models fed by a slightly
differing sequences of pseudo random numbers take ``very
different turns in phase space'' and thus will display very little
correlations.

\subsection{Output}
\slabel{output}
As mentioned above, it is generally advisable to output and flush data
frequently in \keyword{chunks}, resetting accumulating variables
afterwards. Even when output occurs every second, the overhead
in terms of the CPU and real time spent by the system is likely to be
negligibly small.

Where data is written to a file in large quantities or frequently,
\keyword{buffered I/O} as provided by \code{stdio} through the
\code{printf}-family of library calls is usually much faster than
writing immediately to the file using \code{unistd}'s \code{write}.
There are two caveats to this approach: Firstly, depending on the size
of the buffer and thus the frequency of writing, a significant amount of
CPU time may be lost if the program terminates unexpectedly. To avoid
corrupt data, \code{fflush()} should be used rather than allowing the
buffer to empty whenever it reaches its high-water mark. 
Secondly,
if buffering I/O has a significant impact on the computational
performance, the data may better be processed on-the-fly rather than
storing it in a file.

In the following, \code{stdio} is used for its convenient formatting
capabilities, provided by the plethora of flags in the formatting string
of a \code{printf} call. To avoid the problems mentioned above, buffers
are either flushed after each chunk by means of \code{fflush}, or
buffering is switched to buffering line by line, using \code{setlinebuf}.

To avoid unexpected 
interference of the operating system 
with
the simulation, operations should be avoided that can potentially fail because the
environment changes. This applies, in particular, to
read and write access to files. In any case, such operations need to be
encapsulated in an \code{if} condition that catches failing system calls
and triggers a suitable remedy. 

Output of chunks should therefore happen through the \code{stdout}
stream which is by default open at the time of the program start. As the
output is usually used in post-processing it needs to be retained, which
can be achieved by re-directing \code{stdout} into a file. In the
typical shell syntax this can be done in the command line by, say,
\code{./simulation >
output.txt}. To allow easy post-processing, every line should contain
all relevant simulation parameters, such as the system size, the number
of the chunk (a counter), the number of events per chunk, the initial
seed of the \abbrevdef{random number generator}{RNG}, in fact, everything
that is needed to reproduce that line from scratch or to plot the
relevant (derived) data.
Typical examples are
moments to be plotted against the system size and
moment ratios, involving different moments of the same observable.
Using
post-processing tools to wade through vast amounts of data to find the
missing piece of information to amend a line of data can require
significant effort and is highly error-prone. 

Repeating the same output (system size, RNG seed etc) over and over
seemingly goes against
the ethos of avoiding redundant information, which should be
applied when setting up a computer simulation (to avoid clashes), but is
wholly misplaced when it comes to data output. In fact, redundancy in
output is a means to measure consistency and a matter of practicality as
almost all basic post-processing tools are line-oriented.

In some rare cases, an action by the simulation or an event on the
system can result in a \keyword{signal} being sent to the running instance of the
program. In response the program suspends the current operation,
executes a signal handler and continues where it left off. In principle,
the signal should not lead to inconsistent data or behaviour; in fact,
it is probably the most basic but also a very convenient way to communicate
with a running program. For example
\begin{lstlisting}
#include <signal.h>
...
void sighup_handler(int signo);
...
signal(SIGHUP, sighup_handler);
...
void sighup_handler(int signo) 
{
finish_asap=1;
}
\end{lstlisting}
assigns the signal handler \code{sighup\_handler} to deal with the signal
\code{SIGHUP}, which can be sent to the program using \code{kill -HUP}.

There is a rare situation when the signal interrupts in a way that it
leads to unexpected behaviour, namely when it arrives while a ``slow
system call'' is executed, \ie an operation that is performed by the
kernel on behalf of the programme, but which can take a long time to complete, such as
\code{pause}, \code{sleep}, but also \code{write} to so-called pipes.
Without discussing the technical details of the latter, it can lead to
inconsistencies in the output which might not be detected in the
post-processing. For example, a chunk may contain truncated lines and
thus may lack certain information or data,
which the post-processing tools might treat as zeroes. Apart from a
graphical inspection of the data, 
two measures may
therefore be advisable: Firstly, output can be encapsulated in calls of
\code{sigprocmask} which allows temporary suspension of the delivery of
signals. Secondly, a chunk can be terminated by a single line containing
a keyword to indicate the successful completion of the output (\ie
without catching an error, in particular not an ``interrupted system
call'', \code{EINTR}), such as the tag (see
below) \code{\#Completed}. Simply counting the number of occurrences of
that tag and comparing to (supposed) the number of valid chunks can pick
up inconsistencies. In large scale simulations, where disk space can be
a problem leading to truncated files as the system runs out of file
space, this is particularly advisable.

After a chunk has been written out, variables collecting data have to be
reset. Where PDFs are estimated, sweeping across the entire histogram
can become expensive and therefore performing all relevant steps
simultaneously is advantageous for the overall performance. Using one of the
examples above (\Sref{binning}):
\begin{lstlisting}
long long total=0;
for (i=0; i<SMALL2MEDIUM_THRESHOLD; i++) 
  if (histo_small[i]) {
    printf(...);
    total+=histo_small[i];
    histo_small[i]=0;
  }
...
printf("out_of_range: %i\n", histo_out_of_range);
total+=histo_out_of_range;
histo_out_of_range=0;
printf("total: %lli", total);
\end{lstlisting}

The final line allows the user to compare the number of events collected
in the histogram to the number of events expected. It is a
computationally cheap additional check for data consistency.

To distinguish different types of output, such as moments of different
observables, data should be \keyword{tagged} by short keys that are easily
filtered out in post processing. For example, if every line containing moments of
avalanche sizes is tagged by \code{\#M\_SIZE} at the beginning, all relevant lines can be extracted
very easily for example using \code{grep '^#M_SIZE' output.txt}. To
strip off the tags, one either appends \code{|sed 's/\#M\_SIZE//'} or
includes the functionality of \code{grep} in the \code{sed} command,
\begin{lstlisting}
sed -n 's/^#M_SIZE//p' output.txt > output.txt_M_SIZE
\end{lstlisting}
storing all relevant lines in \code{output.txt\_M\_SIZE} for further
processing by other tools. One very simple, but particularly powerful one
is \code{awk}. For example, the average across  the seventh column
starting with the 101st chunk (stored in the first column) can be
calculated using 
\begin{lstlisting}[showstringspaces=false]
awk ' { if ($1>100) {m0++; m1+=$7;} } END { printf ("%i %10.20g\n", m0, m1/m0); } ' output.txt_M_SIZE
\end{lstlisting}
All of this is very easily automated using powerful
scripting languages (in particular shell scripts, \code{awk}, \code{sed} and
\code{grep}), and more powerful (interpreted) programming languages,
such as \code{perl} or \code{python}, which provide easy access to
line-oriented data. In recent years, XML has become more popular to
store simulation parameters as well as simulation results.

\section{Summary and conclusion}
The early life of SOC was all about computer models that showed the
desired features of SOC: Intermittent behaviour (slow drive, fast
relaxation) displaying scale
invariance as observed in traditional critical phenomena without the
need to tune a control parameter to a critical value. After many authors
had (mostly with little success)
attempted to populate the universality class of the BTW Sandpile, a range
of SOC models was proposed firstly as a paradigm of alternative
universality classes and later to highlight specific aspects of SOC, such
as non-conservation (as for example in the Forest-Fire Model),
non-Abelianness
(as for example in the Olami-Feder-Christensen Model) and stochasticity
(as for example in the Manna Model).

Many of these models have been studied extensively, accumulating
hundreds of thousands of hours of CPU time in large-scale Monte Carlo
simulations. A finite size scaling analysis of the data generally
produces a set of two to eight exponents, which are supposedly
universal. It turns out, however, that very few models display
clean, robust scaling behaviour in the event size distribution,
although it is remarkably broad for many models. 

Of the models discussed above, the Manna Model displays the clearest
signs of scale invariance. There is wide consensus that it is the same
universality class as the Oslo Model
\citep{ChristensenETAL:1996,NakanishiSneppen:1997}. In the conservative
limit and in the near-conservative regime, the Olami-Feder-Christensen Model also displays
convincing moment scaling, but less so for smaller values of the level
of conservation. Numerical artefacts may play a significant r{\^o}le in
its scaling \citep{Drossel:2002}.

The Forest Fire Models is widely acknowledged for failing to display
finite size scaling in the event size distribution
\citep{Grassberger:2002a,JensenPruessner:2002b}, although
its moments still display some scaling \citep{JensenPruessner:2004}.
The contrast is even sharper in the Bak-Tang-Wiesenfeld Model:
Some scaling is known analytically
\citep{MajumdarDhar:1992,Ivashkevich:1994,IvashkevichKtitarevPriezzhev:1994a,DharManna:1994}, yet the event size distribution seems at best be
governed by multiscaling
\citep{TebaldiDeMenechStella:1999,Drossel:1999a,Drossel:2000,DornHughesChristensen:2001}

While analytical approaches receive increasing attention, numerical
techniques remain indispensable in the development and analysis of
models which are tailor-made to display specific features or to mimic
experimental systems. Models developed more recently are usually
implemented in C, producing numerical data in
Monte-Carlo simulations. It is fair to say that the careful data
analysis requires as much attention to detail as the implementation of
the model in the first place. 

While the classic data-collapse and more immediate tests for scaling
dominated the early literature of SOC, more recently the finite size
scaling of moments \citep{TebaldiDeMenechStella:1999} has become the
predominant technique for the extraction of scaling exponents. 
Apart from identifying the mechanism of SOC, the main purpose of the
numerical work is to establish universality and
universality classes among models, as well as their relation to natural
phenomena.
One may hope that these efforts will eventually help to uncover the necessary
and sufficient conditions for SOC.

\begin{acknowledgement}
The author gratefully acknowledges the kind support by EPSRC Mathematics Platform
grant EP/I019111/1.
\end{acknowledgement}

\section*{Appendix: Implementation details for binning}
\addcontentsline{toc}{section}{Appendix: Implementation details for
binning}
\slabel{appendix_binning}
To implement binning in computer simulations of SOC models it is
advisable to perform simple bit manipulations on basic,
integer-valued
observables. It often suffices to implemented three levels of coarse
graining or less, for example
\begin{lstlisting}
#define SMALL2MEDIUM_THRESHOLD (1LL<<15)
long long histo_small[SMALL2MEDIUM_THRESHOLD]={0LL};
#define MEDIUM2LARGE_THRESHOLD (1LL<<30)
#define MEDIUM_SHIFT     (12)
long histo_medium[(MEDIUM2LARGE_THRESHOLD-SMALL2MEDIUM_THRESHOLD)>>MEDIUM_SHIFT]={0L};
#define LARGE_THRESHOLD (1LL<<45)
#define LARGE_SHIFT (27)
int histo_large[(LARGE_THRESHOLD-MEDIUM2LARGE_THRESHOLD)>>LARGE_SHIFT]={0};
int histo_out_of_range=0;
long long int s; /* event size */

...

if (s<SMALL2MEDIUM_THRESHOLD) histo_small[s]++;
else if (s<MEDIUM2LARGE_THRESHOLD) histo_medium[(s-SMALL2MEDIUM_THRESHOLD)>>MEDIUM_SHIFT]++;
else if (s<LARGE_THRESHOLD) histo_large[(s-MEDIUM2LARGE_THRESHOLD)>>LARGE_SHIFT]++;
else histo_out_of_range++;
\end{lstlisting}
Here the event size to be tallied is \code{s}. In the block of \code{if}
statements, it is compared to various thresholds before it is rescaled
and counted  into a histogram. Because vectors in many programming
languages start with index $0$, a shift an offset is subtracted as well.
It can pay of to re-arrange the \code{if} statements 
as to test against the most frequent case as early as possible.
One case, in the present
example the last one,
counts the number of times the counter overspills, here
\code{histo\_out\_of\_range}. 

Some subtleties of the above implementation are worth discussing. Firstly, the
types used for the histogram typically decrease in size with increasing
event size while the size
of the type needed to represent the event size at the respective
thresholds increases. This is because normally the frequency is an \emph{inverse}
power law of the event size. Great care must be taken to avoid
unnecessary typecasts and undesired outcomes, as some languages, in
particular C, are rather idiosyncratic when it comes to (integer) type-promotion
in comparisons, in particular when they involve signs. 

In the above examples, automatic vector variables are used and
initialised by assigning \code{\{0\}}, which is expanded by the compiler to a suitable
size by adding zeroes. Initialisation of vectors in C has been further
simplified in the C99 standard.

Secondly,
it is important to choose the thresholds together with
the planned bit-shifts, in order to avoid 
an \emph{off-by-one} error. The problem is that, say,\linebreak
\code{s<MEDIUM2LARGE\_THRESHOLD}, does not imply 
\begin{lstlisting}
(s-SMALL2MEDIUM_THRESHOLD)/((1<<MEDIUM_SHIFT) < (MEDIUM2LARGE_THRESHOLD-SMALL2MEDIUM_THRESHOLD)/(1<<MEDIUM_SHIFT)
\end{lstlisting}
because for some \code{s<MEDIUM2LARGE\_THRESHOLD} their bitshifted
value\linebreak
\code{s>>MEDIUM\_SHIFT} in fact equals
\code{MEDIUM2LARGE\_THRESHOLD>>MEDIUM\_SHIFT}, \linebreak
namely precisely when
\code{MEDIUM2LARGE\_THRESHOLD} is not an integer multiple of
\code{1<<MEDIUM\_SHIFT}. It is therefore a matter of defencive
programming to write the thresholds for the macros in this form:
\begin{lstlisting}
#define MEDIUM2LARGE_THRESHOLD ((1LL<<18) * (1LL<<MEDIUM_SHIFT))
\end{lstlisting}

As for a rudimentary output routine 
\begin{lstlisting}
for (i=0; i<SMALL2MEDIUM_THRESHOLD; i++) 
  if (histo_small[i]) printf("%i %i %lli %i\n", i, i, histo_small[i], 1);
for (i=0; i<((MEDIUM2LARGE_THRESHOLD-SMALL2MEDIUM_THRESHOLD)>>MEDIUM_SHIFT); i++) 
  if (histo_medium[i]) printf("%li %i %li %i\n", ((long)SMALL2MEDIUM_THRESHOLD)+(((long)(i))<<MEDIUM_SHIFT), i, histo_medium[i], 1<<MEDIUM_SHIFT);
for (i=0; i<((LARGE_THRESHOLD-MEDIUM2LARGE_THRESHOLD)>>LARGE_SHIFT); i++) 
  if (histo_large[i]) printf("%lli %i %i %i\n", ((long long)MEDIUM2LARGE_THRESHOLD)+(((long long)(i))<<LARGE_SHIFT), i, histo_large[i], 1<<LARGE_SHIFT);
printf("out_of_range: %i\n", histo_out_of_range);
\end{lstlisting}
care must again be taken that the formatting of the output is in line
with the type of the data and does not
spoil it. Fortunately, most modern compilers spot clashes between the
formatting string used in \code{printf}
and the actual argument. As discussed below, it is generally advisable
to have only one output stream, namely \code{stdout}, and to use
\keyword{tags} to mark up data for easy fetching by
post-processing tools. In the example above, the bins have not been
rescaled by their size which instead has been included explicitly in the
output. A sample of the PDF can be inspected by plotting the third
column divided by the fourth against the first.

\newcommand{\bibconferencename}[1]{\emph{#1}}

\bibliography{chapter_soc_simulations}

\end{document}